\documentclass[twocolumn]{aastex62}
\hypersetup{linkcolor=magenta,citecolor=blue,filecolor=cyan,urlcolor=purple}

\usepackage{graphicx}
\usepackage[caption=false]{subfig}
\usepackage{ulem}

\received{January 24, 2019}
\revised{March 26, 2019}
\accepted{April 9, 2019}
\submitjournal{ApJ}

\shorttitle{Multipoint Study of Successive CMEs}
\shortauthors{Palmerio et al.}

\turnoffediting

\begin{document}

\title{Multipoint Study of Successive Coronal Mass Ejections Driving Moderate Disturbances at 1 AU}

\correspondingauthor{Erika Palmerio}
\email{erika.palmerio@helsinki.fi}

\author[0000-0001-6590-3479]{Erika Palmerio}
\affil{Department of Physics, University of Helsinki, P.O. Box 64, FI-00014 Helsinki, Finland}

\author[0000-0002-5681-0526]{Camilla Scolini}
\affil{Centre for mathematical Plasma Astrophysics (CmPA), KU Leuven, 3001 Leuven, Belgium}
\affil{Solar--Terrestrial Centre of Excellence---SIDC, Royal Observatory of Belgium, 1180 Brussels, Belgium}

\author[0000-0003-1137-8220]{David Barnes}
\affil{STFC-RAL Space, Rutherford Appleton Laboratory, Harwell Campus, OX11 0QX, UK}

\author[0000-0003-1169-3722]{Jasmina Magdaleni{\'c}}
\affil{Solar--Terrestrial Centre of Excellence---SIDC, Royal Observatory of Belgium, 1180 Brussels, Belgium}

\author[0000-0002-0631-2393]{Matthew J. West}
\affil{Solar--Terrestrial Centre of Excellence---SIDC, Royal Observatory of Belgium, 1180 Brussels, Belgium}

\author[0000-0002-2542-9810]{Andrei N. Zhukov}
\affil{Solar--Terrestrial Centre of Excellence---SIDC, Royal Observatory of Belgium, 1180 Brussels, Belgium}
\affil{Skobeltsyn Institute of Nuclear Physics, Moscow State University, 119991 Moscow, Russia}

\author[0000-0002-6097-374X]{Luciano Rodriguez}
\affil{Solar--Terrestrial Centre of Excellence---SIDC, Royal Observatory of Belgium, 1180 Brussels, Belgium}

\author[0000-0003-4105-7364]{Marilena Mierla}
\affil{Solar--Terrestrial Centre of Excellence---SIDC, Royal Observatory of Belgium, 1180 Brussels, Belgium}
\affil{Institute of Geodynamics of the Romanian Academy, 020032 Bucharest-37, Romania}

\author[0000-0002-4921-4208]{Simon W. Good}
\affil{Department of Physics, University of Helsinki, P.O. Box 64, 00014 Helsinki, Finland}

\author[0000-0002-8416-1375]{Diana E. Morosan}
\affil{Department of Physics, University of Helsinki, P.O. Box 64, 00014 Helsinki, Finland}

\author[0000-0002-4489-8073]{Emilia K. J. Kilpua}
\affil{Department of Physics, University of Helsinki, P.O. Box 64, 00014 Helsinki, Finland}

\author[0000-0003-1175-7124]{Jens Pomoell}
\affil{Department of Physics, University of Helsinki, P.O. Box 64, 00014 Helsinki, Finland}

\author[0000-0002-1743-0651]{Stefaan Poedts}
\affil{Centre for mathematical Plasma Astrophysics (CmPA), KU Leuven, 3001 Leuven, Belgium}



\begin{abstract}
We analyse in this work the propagation and geoeffectiveness of four successive coronal mass ejections (CMEs) that erupted from the Sun during 21--23 May 2013 \edit1{and that were detected in interplanetary space by the \emph{Wind} and/or STEREO-A spacecraft. All these CMEs featured critical aspects for understanding so-called ``problem space weather storms'' at Earth. In the first three events a limb CMEs resulted in moderately geoeffective in-situ structures at their target location in terms of the disturbance storm time (\emph{Dst}) index (either measured or estimated). The fourth CME, which also caused a moderate geomagnetic response, erupted from close to the disc centre as seen from Earth, but it was not visible in coronagraph images from the spacecraft along the Sun--Earth line and appeared narrow and faint from off-angle viewpoints. Making the correct connection between CMEs at the Sun and their in-situ counterparts is often difficult for problem storms. We investigate these four CMEs using multiwavelength and multipoint remote-sensing observations (extreme ultraviolet, white light, and radio), aided by 3D heliospheric modelling, in order to follow their propagation in the corona and in interplanetary space and to assess their impact at 1~AU. Finally, we emphasise the difficulties in forecasting moderate space weather effects provoked by problematic and ambiguous events and the importance of multispacecraft data for observing and modelling problem storms.}
\end{abstract}

\keywords{Sun: coronal mass ejections (CMEs) -- solar--terrestrial relations -- solar wind}


\section{Introduction} \label{sec:intro}
Coronal mass ejections \citep[CMEs; e.g.,][]{webb2012} are well known to be the principal drivers of space weather effects at Earth \citep[e.g.,][]{gosling1991,gonzalez1999,huttunen2005,koskinen2006,richardson2012}. The subset of CMEs that are most likely to drive geomagnetic disturbances are front-sided full halos that are seen to entirely encompass the solar disc in the field of view of coronagraphs along the Sun--Earth line \citep[e.g.,][]{howard1982,webb2000,srivastava2004,schwenn2005,gopalswamy2007,zhang2007,scolini2018a}. Another important but less accurately predictable (in terms of hit/miss) subset of CMEs are partial halos, which are seen to erupt with a wide angle in coronagraph images without forming a complete ring around the solar disc.

The source region of a halo CME can be located anywhere on the solar disc, but it has been shown that the most geoeffective ones tend to originate closer to the central meridian \citep{gopalswamy2007}. Nevertheless, some limb halo CMEs (i.e., source region located $> \pm 45^{\circ}$ longitude from the central meridian) have been observed to drive geomagnetic storms \citep{huttunen2002,rodriguez2009,gopalswamy2010b,cid2012}. \citet{cid2012} studied 25 full halo CMEs that erupted from the limb during solar cycle 23, and concluded that four of them (all coming from the West limb) were drivers of geomagnetic activity. This suggests that limb CMEs, and in particular limb halos, should be taken into consideration in space weather predictions, although as sources of moderate disturbances only since they usually make glancing encounters with Earth.

However, the geoeffectiveness of limb halos is usually more difficult to predict than for disc halos (i.e., source region located $< \pm 45^{\circ}$ longitude from the central meridian). In general, all CMEs are affected by a certain degree of unpredictability as they travel away from the Sun, mostly because of deflections \citep[e.g.,][]{wang2014,kay2015,kay2016}, rotations \citep[e.g.,][]{mostl2008,yurchyshyn2009,vourlidas2011,isavnin2014}, deformations \citep[e.g.,][]{savani2010}, and/or interactions with other CMEs or other heliospheric structures \citep{lugaz2012,lugaz2017,shen2012}. In the case of limb halos, it is particularly uncertain whether a CME will hit Earth at all. Another aspect to take into account is that although the ejecta of a limb halo may miss Earth, the related interplanetary shock and sheath may instead result in an impact. CME-driven sheaths are well-known drivers of significant geomagnetic disturbances \citep[e.g.,][]{tsurutani1988,gonzalez1999,gonzalez2011,huttunen2002,huttunen2004,lugaz2016,kilpua2017a}. \citet{gopalswamy2010b} studied 17 limb halos and their interplanetary counterparts, and concluded that the geoeffectiveness was caused by the sheath region in all the cases in which the association could be made unambiguously.

Another class of CMEs for which it is difficult to assess geoeffectiveness is represented by those CMEs that are not visible in coronagraph imagery because they are too narrow and/or too faint \citep[e.g.,][]{yashiro2005,vourlidas2017}. Such CMEs are mostly missed when viewed along the Sun--spacecraft line, but there are cases when CMEs are extremely faint in coronagraph observations even from an off-angle view \citep[e.g.,][]{kilpua2014}. \edit1{Furthermore, \citet{howard2008b} identified a significant number of}\ CME events observed by the \textit{Solar Mass Ejection Imager} \citep[SMEI;][]{eyles2003} in the inner heliosphere that were not visible in coronagraph data. This suggests that these events contained initially little to no excess mass compared to the ambient coronal density, but gained mass during their propagation in the inner heliosphere. Assessing the geoeffectiveness of such events from a single viewpoint would be a highly challenging task, since little to nothing could be said about their propagation speed or direction. Furthermore, \citet{schwenn2005} accounted in a statistical study that spanned four years of data that about 20\% of interplanetary CMEs \citep[or ICMEs; e.g.,][]{kilpua2017b} and related storms did not have a front-sided \edit1{halo (partial or full)}\ CME source.

In this article, we further investigate and discuss the issue of observing and forecasting ``problematic'' CMEs. We study a series of four CMEs that erupted during 21--23 May 2013 and whose associated interplanetary shocks, sheaths, and ejecta reached Earth and the STEREO-A spacecraft. The CMEs studied can all be considered as ``problematic'' from a forecasting perspective because they either originated from the solar limb with respect to their target location or they were not visible (or extremely faint) in coronagraph imagery. Nevertheless, all CMEs that arrived at Earth caused moderate geomagnetic activity in terms of the disturbance storm time (\emph{Dst}) index. At STEREO-A, we evaluate the ``geoeffectiveness'' of the observed CMEs using existing \emph{Dst} prediction formulas that take solar wind parameters as input. We investigate in particular whether these CMEs are observed in multiwavelength and multipoint remote-sensing observations, including extreme ultraviolet (EUV), radio, and white-light coronagraph and heliospheric imager data, and to what extent techniques based on these observations can predict the impact and arrival time of the CMEs. This analysis is complemented by performing a simulation using the 3D heliospheric model EUHFORIA. In Section~\ref{sec:data}, we introduce the spacecraft and the instruments that we use in this work. In Section~\ref{sec:overview}, we present a complete observational overview or the CMEs under study. In Section~\ref{sec:analysis}, we present a detailed analysis of these events, from both the observational and the modelling perspectives. Finally, in Section~\ref{sec:conclusions} we discuss and summarise our results.

\section{Spacecraft Data} \label{sec:data}

The solar disc from Earth's view is imaged by the \textit{Solar Dynamics Observatory} \citep[SDO;][]{pesnell2012} and the \textit{Project for On Board Autonomy 2} \citep[PROBA2;][]{santandrea2013}. 
Line-of-sight photospheric magnetograms are provided by the \textit{Helioseismic and Magnetic Imager} \citep[HMI;][]{scherrer2012} onboard SDO. EUV observations are provided by both the \textit{Atmospheric Imaging Assembly} \citep[AIA;][]{lemen2012} onboard SDO and the \textit{Sun-Watcher with Active Pixel System and Image Processing} \citep[SWAP;][]{halain2013,seaton2013} onboard PROBA2. The combination of the two different EUV instruments enables us to observe the Sun in several SDO/AIA channels and to take advantage of the enlarged field of view of PROBA2/SWAP that is especially useful for observing CMEs off limb. 

Solar observations from other viewpoints are made with the \textit{Sun Earth Connection Coronal and Heliospheric Investigation} \citep[SECCHI;][]{howard2008a} \textit{Extreme UltraViolet Imager} (EUVI) onboard the \textit{Solar Terrestrial Relations Observatory} \citep[STEREO;][]{kaiser2008}. The STEREO mission consists of twin spacecraft that orbit the Sun, one ahead of Earth in its orbit (STEREO-A) and the other trailing behind (STEREO-B).

After the onset of eruptions, we follow the evolution of the CMEs through coronagraphs and heliospheric imagers. Coronagraph observations are made from three vantage points. The view from Earth is provided by the \textit{Large Angle and Spectrometric Coronagraph} \citep[LASCO;][]{brueckner1995} C2 and C3 instruments onboard the \textit{Solar and Heliospheric Observatory} \citep[SOHO;][]{domingo1995}. The views from STEREO-A and STEREO-B are provided by the SECCHI COR1 and COR2 coronagraphs. We observe the space between the outer corona ($\sim 15$~R$_{\odot}$) and Earth through the \textit{Heliospheric Imagers} \citep[HI;][]{eyles2009} onboard the twin STEREO spacecraft. Each HI instrument comprises two cameras, HI1 and HI2.

In-situ measurements from Earth's Lagrange L1 point are taken with the \textit{Wind} \citep{ogilvie1997} satellite. We use data from the \textit{Magnetic Field Investigation} \citep[MFI;][]{lepping1995}, the \textit{Solar Wind Experiment} \citep[SWE;][]{ogilvie1995}, and the \textit{Radio and Plasma Wave Investigation} \citep[W/WAVES;][]{bougeret1995} instruments.

Measurements at the STEREO spacecraft are taken with the in-situ instruments \textit{In situ Measurements of Particles And CME Transients} \citep[IMPACT;][]{luhmann2008}, \textit{Plasma and Suprathermal Ion Composition} \citep[PLASTIC;][]{galvin2008}, and \textit{Radio and Plasma Wave Investigation} \citep[S/WAVES;][]{bougeret2008}.

\section{Overview of the 21--23 May 2013 CMEs} \label{sec:overview}

\begin{figure}[ht!]
\epsscale{0.9}
\plotone{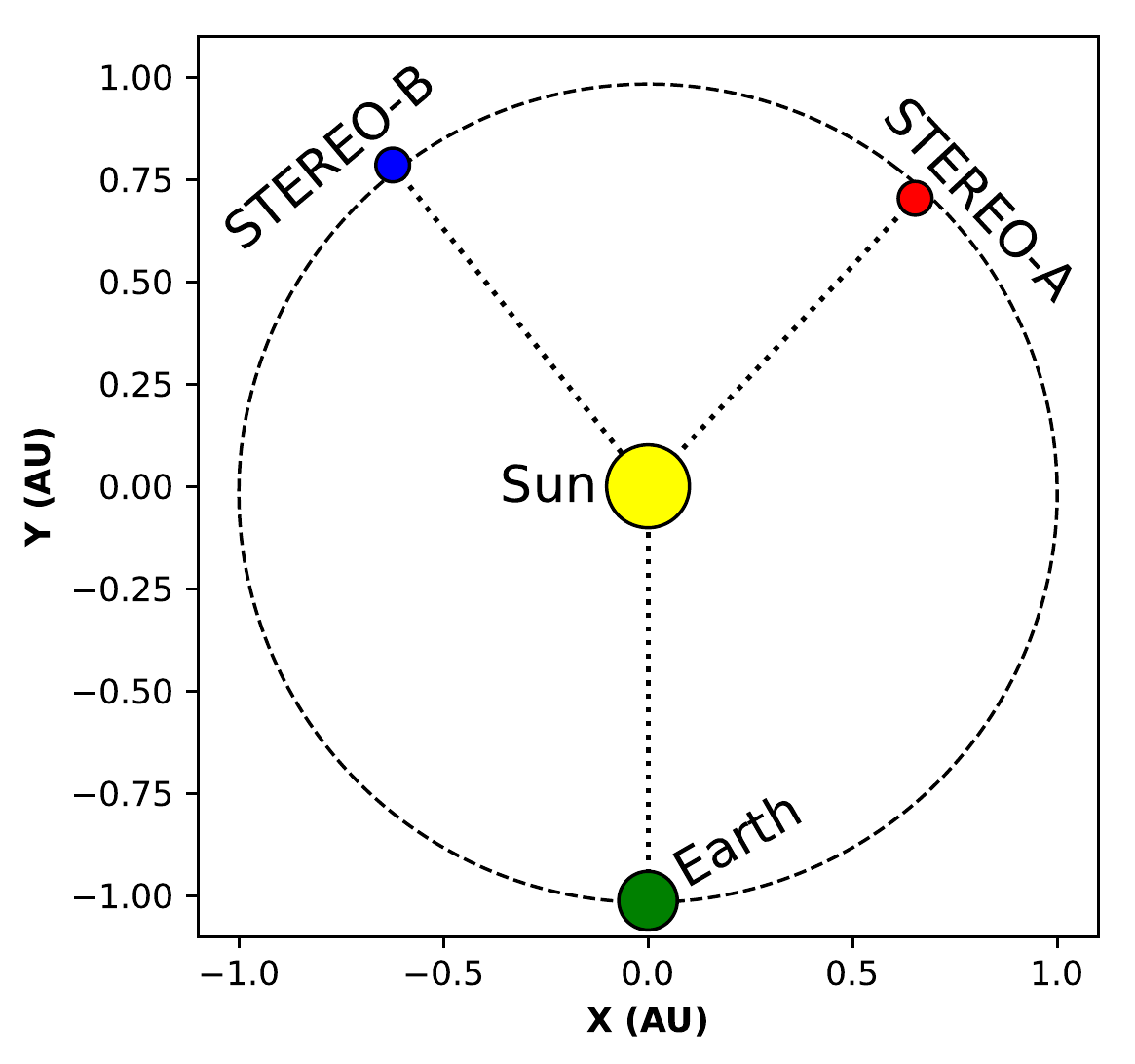}
\caption{The position of Earth and the twin STEREO spacecraft on 22 May 2013, 12:00~UT. The longitudinal separation was $137^{\circ}$ between Earth and STEREO-A, $141^{\circ}$ between Earth and STEREO-B, and $82^{\circ}$ between STEREO-A and STEREO-B. \label{fig:position}}
\end{figure}

We analyse in this study four CMEs that erupted between 21--23~May~2013. Figure~\ref{fig:position} shows the configuration of Earth and the twin STEREO spacecraft roughly in the middle of the selected observation period, i.e.\ on 22~May, 12:00~UT.

\subsection{Remote-sensing Observations} \label{subsec:remotesensing}

\begin{figure*}
\epsscale{1.15}
\plotone{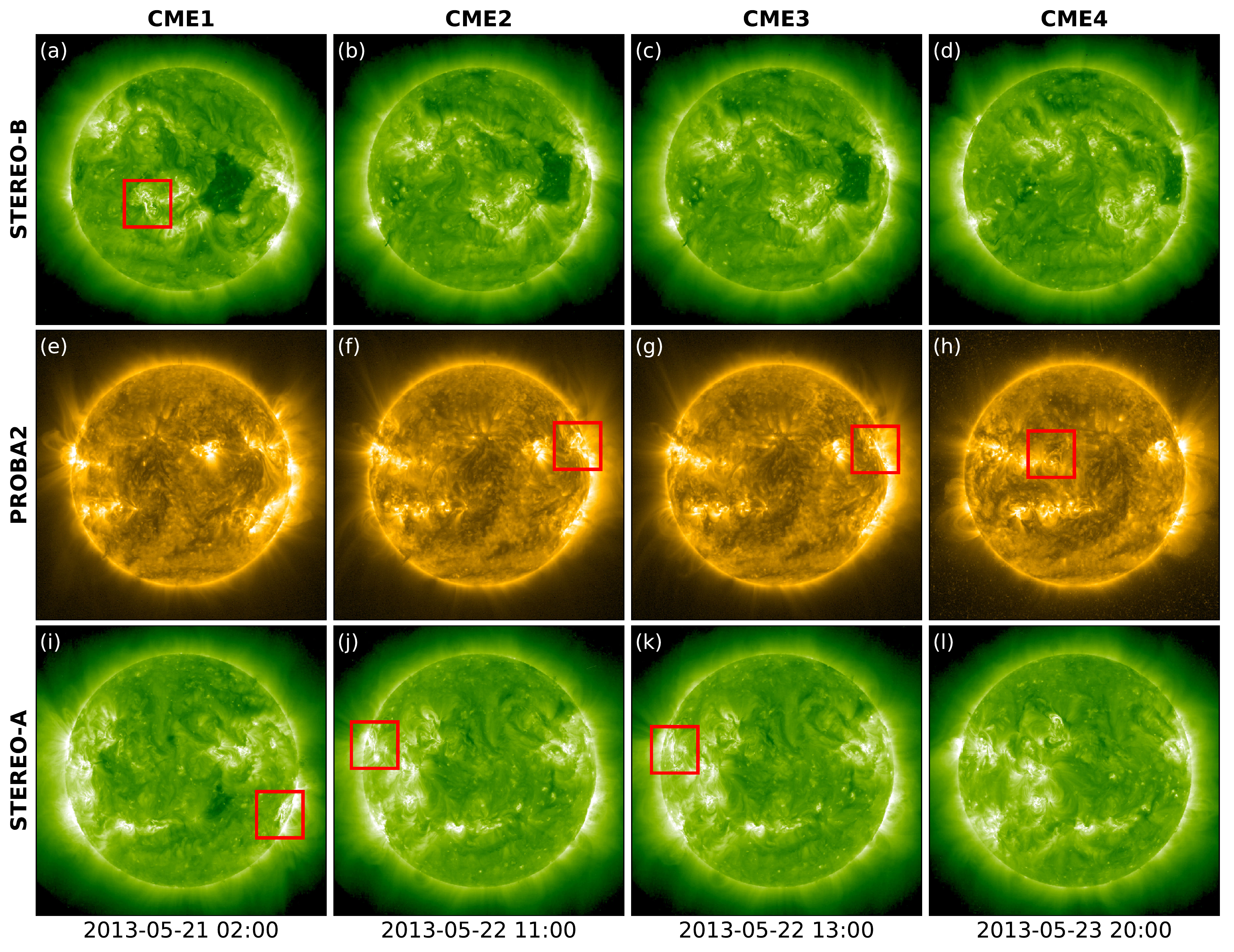}
\caption{Full-disc images from PROBA2/SWAP and STEREO/SECCHI/EUVI around the time of the four CMEs under study. (a--d, top row) Images from STEREO/SECCHI/EUVI-B. (e--h, middle row) Images from PROBA2/SWAP. (i--l, bottom row) Images from STEREO/SECCHI/EUVI-A. The locations of the source regions are indicated with red squares in the panels when they were visible on the disc.\label{fig:fulldisc}}
\end{figure*}

\begin{figure*}
\epsscale{1.15}
\plotone{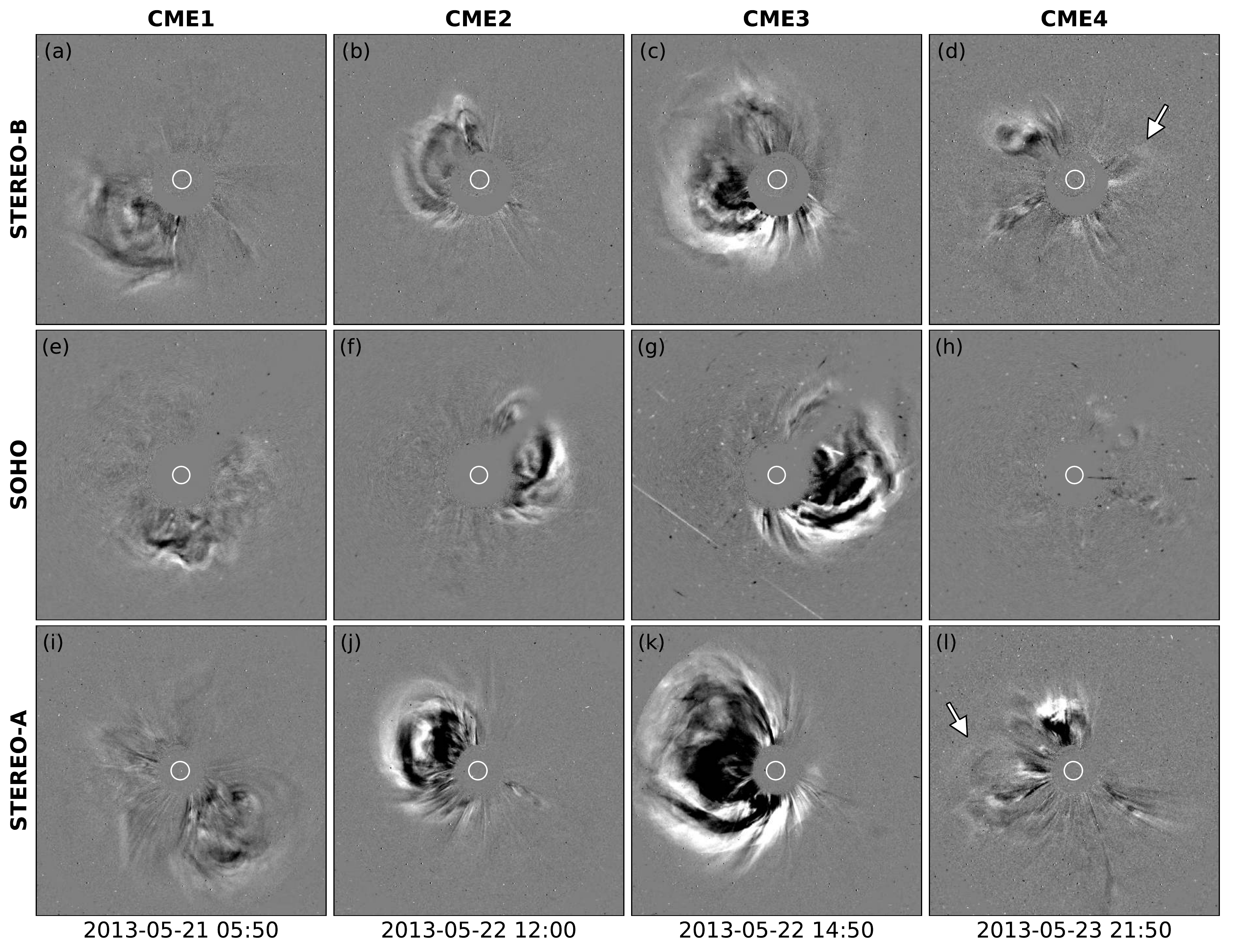}
\caption{The four CMEs under study as seen in running-difference coronagraph images from STEREO/SECCHI/COR2-B (a--d, top row), SOHO/LASCO/C3 (e--h, middle row), and STEREO/SECCHI/COR2-A (i--l, bottom row). CME4 was the faintest of all and is indicated with an arrow in the COR2 images, whilst it was not visible in LASCO data. For each column, all images are taken within eight minutes around the reported time. \label{fig:coronagraph}}
\end{figure*}

\begin{figure*}
\epsscale{1.15}
\plotone{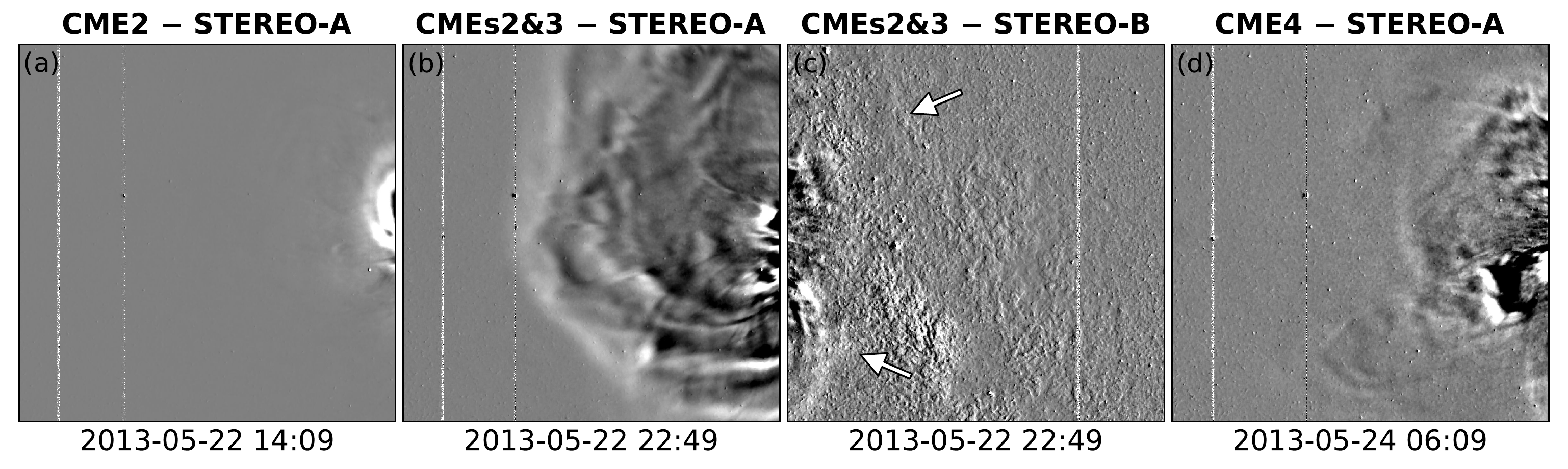}
\caption{Three of the four CMEs under study as seen in running-difference STEREO/SECCHI/HI1 images. The observed CME and the observing spacecraft (A or B) are indicated on top of each image. CME1 was not visible from either of the STEREO spacecraft, and STEREO-B could only detect the large, merged CMEs2\&3 due to the presence of the galactic plane in its field of view (the CME front is indicated with arrows). \label{fig:hi}}
\end{figure*}

The onset and lower coronal signatures of each of the eruptions are revealed from different perspectives in EUV observations from SWAP onboard PROBA2 and EUVI from the SECCHI suite on the STEREO satellites. Figure~\ref{fig:fulldisc} summarises the positions of the source regions of all the CMEs under study on the different spacecraft. After studying the different eruptions on disc, we follow them in coronagraph imagery provided by the SOHO and STEREO satellites. Figure~\ref{fig:coronagraph} shows simultaneous snapshots of each CME from the three viewpoints. \edit1{Finally, we search for signatures of the CMEs in the HI cameras. Figure~\ref{fig:hi} shows snapshots of the CMEs that we could identify in HI1 imagery.}

The first CME (hereafter CME1) erupted on 21~May 2013, $\sim$01:00~UT. As shown in the first column of Figure~\ref{fig:fulldisc}, the CME originated from the SW limb from STEREO-A's viewpoint and from close to the central meridian (in the SE quadrant) from STEREO-B's viewpoint. The source region was located at $(\theta,\phi)=(-20^{\circ},-155^{\circ})$ in Stonyhurst coordinates \citep{thompson2006}, and the eruption was back-sided with respect to the Sun--Earth line (i.e., its source region was not in view for PROBA2/SWAP). This means that its corresponding source region was not classified, but was labelled as NOAA Active Region (AR) 11758 when it rotated onto the Earth-facing solar disc. Upon eruption, coronal dimmings could be seen to extend mostly towards the Southeast in images from both STEREO-A and STEREO-B (we, however, remark that at STEREO-A this may be due to projection effects). This gives a first-order indication that the CME was launched non-radially towards STEREO-A and away from STEREO-B, since the locations of coronal dimmings are believed to generally map to the CME extent in coronagraphs \citep[e.g.,][]{thompson2000}. Such a non-radial propagation may be explained by the presence of a large coronal hole to the west of the source region \citep[e.g.,][]{cremades2004,gopalswamy2009}. Coronagraph data (first column of Figure \ref{fig:coronagraph}) are consistent with this assumption, showing that the main body of CME1 propagated towards the southeast in STEREO-B's field of view despite originating from close to the disc centre. Since this CME was back-sided with respect to Earth, it was not detected in the fields of view of the HI instruments.

The second CME (hereafter CME2) erupted on 22~May~2013, $\sim$07:00~UT. As shown in the second column of Figure~\ref{fig:fulldisc}, the CME originated from the NW limb from Earth's viewpoint and from the NE limb from STEREO-A's viewpoint. The source region was classified as NOAA AR 11745 and was located at $(\theta,\phi)=(15^{\circ},65^{\circ})$ in Stonyhurst coordinates. This indicates that the eruption was back-sided with respect to STEREO-B's viewpoint. This CME erupted from higher up in the solar atmosphere, and a series of erupting loops could be seen off limb in images from both PROBA2 and STEREO-A. The loops reached an altitude of $\sim1.4\,R_{\odot}$ (in plane-of-sky images) shortly before the CME onset, and the only eruption signature that could be seen on disc from both satellites was a set of post-eruption arcades (PEAs). As seen from coronagraph data (second column of Figure~\ref{fig:coronagraph}), this CME appeared as a partial halo from all three viewpoints.

The third CME (hereafter CME3) erupted on 22~May~2013, $\sim$12:30~UT, from the same source region as CME2, located at $(\theta,\phi)=(13^{\circ},70^{\circ})$ in Stonyhurst coordinates. The third column of Figure~\ref{fig:fulldisc} shows that, as CME2, this CME erupted from the NW limb from Earth's viewpoint and from the NE limb from STEREO-A's viewpoint, and was again back-sided with respect to STEREO-B's viewpoint. The source of this eruption was a reverse-S shaped filament (seen in 304~{\AA} data from SDO/AIA and STEREO/SECCHI/EUVI-A) that was lying under the PEAs that originated from CME2. Coronagraph data (third column of Figure \ref{fig:coronagraph}) reveal that this CME appeared as a full halo from all three viewpoints, and that it was significantly faster than CME2. In the LASCO/C3 coronagraph onboard SOHO, it can be seen that CME3 caught up with CME2 (and possibly merged) around 15:30~UT at a plane-of-sky heliocentric distance of $\sim20\,R_{\odot}$. As a further confirmation of this, CME2 was observed in two images in HI1-A (Figure~\ref{fig:hi}(a)) before it was subsumed by CME3 (Figure~\ref{fig:hi}(b)). In the HI1-B field of view, CME2 and CME3 were only visible as a single CME (hereafter CMEs2\&3, Figure~\ref{fig:hi}(c)).

The fourth CME (hereafter CME4) erupted on 23~May~2013, $\sim$19:30~UT, from close to the disc centre (in the NE quadrant) from Earth's viewpoint. The fourth column of Figure~\ref{fig:fulldisc} shows the source region of this CME. Because of the configuration of the spacecraft at this time, the source region could not be observed simultaneously in any of the STEREO satellites, being close to the central meridian on PROBA2. The source region was located at $(\theta,\phi)=(7^{\circ},-13^{\circ})$ in Stonyhurst coordinates, and the eruption originated from between the western edge of NOAA AR 11753 and an adjacent region of more diffuse field. A filament (seen in SDO/AIA 304~{\AA} images) was running along the polarity inversion line (PIL) between these two regions, and the CME originated from its eruption. The on-disc observations of this event show several eruption-associated signatures \citep[e.g.,][]{hudson2001,zhukov2007}: ejection of filament material, flare ribbons, coronal dimmings, and PEAs. The situation, however, is different in the coronagraph data (last column of Figure~\ref{fig:coronagraph}), where the CME appeared faint to invisible, depending on the viewpoint. Only in the COR2-A coronagraph was CME4 visible as a three-part structure \citep{illing1985} along the ecliptic plane, although relatively narrow and faint. In the COR2-B coronagraph, only a very faint jet-like emission could be seen. Finally, this CME was not visible at all on either of the LASCO coronagraphs or the COR1 coronagraphs. Despite being faint and narrow, CME4 appeared to be fairly fast, with a plane-of-sky speed of about 1200~km s$^{-1}$ (in the plane of sky of STEREO-A). Finally, this CME was also visible in the HI field of view, although from STEREO-A only, again as a narrow CME that propagated mostly along the ecliptic plane \edit1{(Figure~\ref{fig:hi}(d))}.

\subsection{\textit{In-situ} Observations} \label{subsec:insitu}

\begin{figure*}
\gridline{\fig{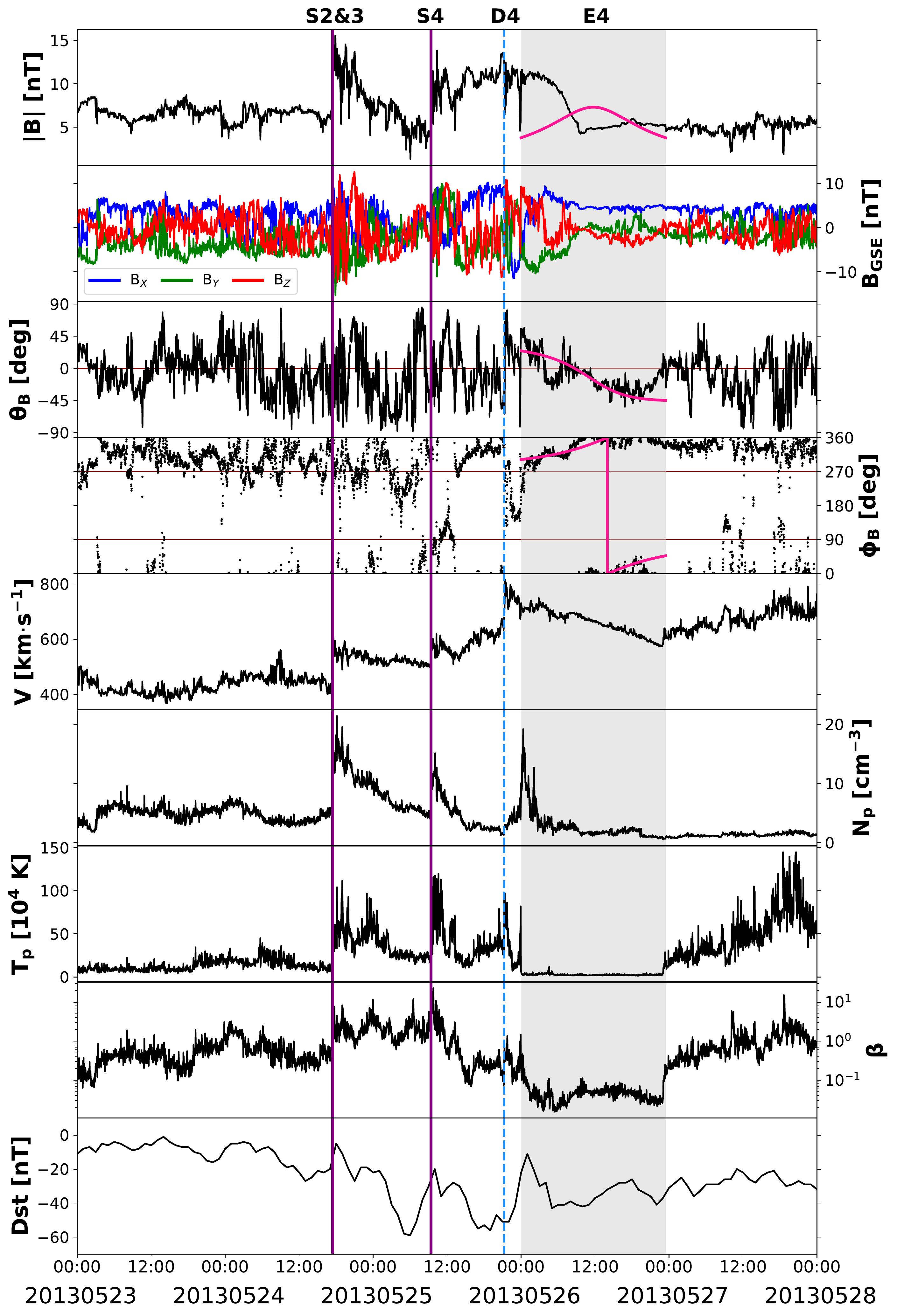}{0.4525\textwidth}{(a) \textit{Wind}}
         \fig{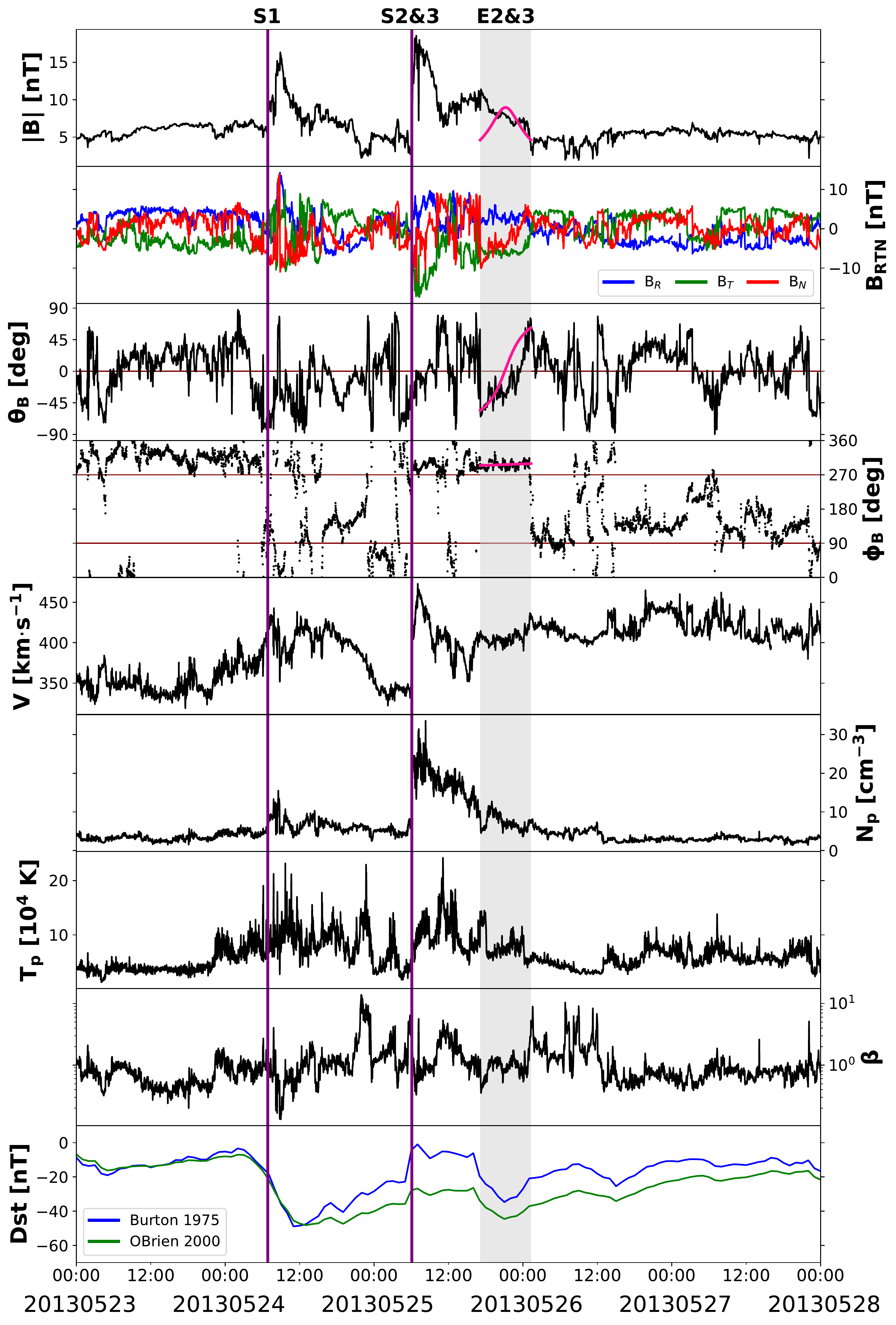}{0.45\textwidth}{(b) STEREO-A}}
\caption{In-situ observations from the (a) \textit{Wind} and (b) STEREO-A spacecraft. Both plots show the following parameters: magnetic field magnitude, Cartesian components of the magnetic field, $\theta$ and $\phi$ (angular) components of the magnetic field, solar wind speed, proton density, proton temperature, plasma $\beta$, and \emph{Dst} index. The purple solid lines represent the shock arrivals (S1, S2\&3, and S4), the cyan dashed line marks the discontinuity in the speed profile (D4), and the grey shaded regions highlight the ICME ejecta (E2\&3 and E4). A Gold--Hoyle flux rope fitting has been overplotted in pink within the ejecta intervals. \label{fig:insitu}}
\end{figure*}

Figure \ref{fig:insitu} shows in-situ measurements taken at \textit{Wind} and STEREO-A during the days following the eruptions described in Section~\ref{subsec:remotesensing}. According to our remote-sensing analysis, we estimate that the \edit1{interplanetary shocks driven by the} CMEs under study arrived at 1~AU in the following way: the shock driven by CME1 \edit1{(hereafter S1)} arrived at STEREO-A on 24~May \edit1{at 06:52~UT}, the shock driven by the merged CMEs2\&3 \edit1{(hereafter S2\&3)} arrived at \textit{Wind} on 24~May \edit1{at 17:26~UT} and at STEREO-A on 25~May \edit1{at 06:05~UT}, and finally the shock driven by CME4 \edit1{(hereafter S4)} arrived at \textit{Wind} on 25~May \edit1{at 09:22~UT}. There are no signatures of CME1 in the STEREO-B in-situ data (not shown), suggesting that CME1 was indeed \edit1{deflected} towards STEREO-A as suggested by the EUV and white-light imaging observations. The other three CMEs were also not observed at STEREO-B.

Since Earth and STEREO-A were separated by about $137^{\circ}$ in longitude (see Figure \ref{fig:position}) during the period under study, we investigate the shock normals (taken from the Heliospheric Shock Database) and the timing between the spacecraft to ensure that S2\&3 was indeed observed at both locations. Concerning the shock normals, we estimate the angle $\alpha$ between the shock normal and the radial direction. The sign of $\alpha$ indicates whether the shock was encountered towards the East or West from its nose. At \emph{Wind}, we obtain $\alpha=-38^{\circ}$, suggesting an encounter towards the East and at STEREO-A, we obtain $\alpha=9^{\circ}$, suggesting an encounter towards the West (here, the plus sign is defined towards the West with respect to the radial direction). The solar wind speed before the shock at \emph{Wind} was $V\sim450$~km$\cdot$s$^{-1}$, and at STEREO-A $V\sim340$~km$\cdot$s$^{-1}$, whilst after the shock passage, the measured speeds are $V\sim550$~km$\cdot$s$^{-1}$ and $V\sim440$~km$\cdot$s$^{-1}$, respectively. At both locations, the speed jump at the shock is thus $\Delta V\simeq100$~km$\cdot$s$^{-1}$. Together with the fact that the shock was detected at STEREO-A about 12~hours later than at \emph{Wind} because of the slower background wind, this suggests that the two spacecraft likely detected the same shock. Our heliospheric simulation constrained by coronagraph observations and presented in Section~\ref{subsec:euhforia} gives also further support that the interplanetary shock driven by CMEs2\&3 was observed at both \emph{Wind} and STEREO-A.

Furthermore, we identify in in-situ data two relatively weak structures that show some clear ICME ejecta signatures \citep[for a description of ICME signatures see, e.g.,][]{zurbuchen2006,kilpua2017b}. These periods are shown in Figure~\ref{fig:insitu} within grey shaded areas. These ejecta would thus correspond to CME4 (hereafter E4, at \emph{Wind}) and to the merged CMEs2\&3 (hereafter E2\&3, at STEREO-A). In order to corroborate these identifications, we look for these two events in the existing ICME catalogues \citep{richardson2010,jian2018,nieveschinchilla2018}. We find that E4, observed at \emph{Wind}, is reported in the Richardson~\&~Cane catalogue, whilst E2\&3, observed at STEREO-A, is reported in the Jian et al. catalogue. Neither ejecta, however, has been classified as a magnetic cloud, i.e., a structure showing enhanced magnetic field magnitude, a smoothly rotating magnetic field over one direction, and low plasma temperature and beta \citep{burlaga1981}. E4 at \emph{Wind} exhibits low plasma temperature and beta, and the magnetic field shows a modest rotation. In turn, E2\&3 at STEREO-A does not show depressed temperature nor plasma beta, but the magnetic field exhibits a clear rotation. Both ejecta display similar magnetic field magnitudes, with peak values of $\sim10$~nT.

In order to further confirm the CME--ICME associations described above, we compare the helicity sign of the corresponding flux ropes at the Sun \edit1{with the magnetic structures in situ}. At the Sun, we use proxies from multiwavelength observations \citep[e.g.,][]{palmerio2017,palmerio2018}; at 1~AU, we apply the Gold--Hoyle flux rope fitting technique \citep{gold1960,farrugia1999}, shown in Figure~\ref{fig:insitu} at both spacecraft. We note clear reverse-J flare ribbons after the eruption of both CME3 and CME4, which are a sign of negative helicity. This is consistent with the helicity sign given by the flux rope fits to the in-situ data, which is also negative in both cases.

Additionally, we note at \emph{Wind} a substantial increase in the solar wind speed from $\sim620$~km$\cdot$s$^{-1}$ to $\sim800$~km$\cdot$s$^{-1}$ on 24~May in the interval 21:10--21:20~UT (marked as D4 in Figure~\ref{fig:insitu}(a)), i.e., occurring within the sheath between S4 and E4, close to the E4 leading edge. At this time, the solar wind density and temperature show a slight increase, the $B_{Z}$ component of the magnetic field rotates abruptly from the South to the North, whilst the magnetic field magnitude features a dip, resembling a magnetic hole \citep[e.g.,][]{turner1977}. These observations, together with the fact that E4 shows a faster speed than its corresponding shock at 1~AU, suggest significant interaction of S4 with the ambient solar wind. Namely, it is likely that S4 was initially travelling at a faster speed through the medium-to-fast stream that immediately follows E4 at \emph{Wind}, but successively slowed down when the shock wave encountered the slow stream ahead of it. In such a scenario, D4 would represent the interface between the faster and slower ambient streams. Previous studies have shown that the propagation direction and speed of a CME ejecta is more constrained by the structure of the background wind than its corresponding shock, which is able to expand through different ambient streams \citep[e.g.,][]{wood2012}. The faster stream that follows E4 ($V\sim600$--$700$~km$\cdot$s$^{-1}$) likely originated from the extended coronal hole to the North of the CME4 source region, visible in Figure~\ref{fig:fulldisc}(h).

Finally, in order to evaluate the geomagnetic impact of all the CMEs, we analyse the \emph{Dst} index profiles. For those CMEs that reached Earth,  we use hourly \emph{Dst} values from the World Data Center for Geomagnetism, Kyoto. At STEREO-A, we calculate the \emph{Dst} index from the solar wind data using the models by \citet{burton1975} and \citet{obrien2000}. We initially use 1-minute solar wind data and then resample the calculated \emph{Dst} to 1-hour cadence. The results are shown in the bottom panels of Figure \ref{fig:insitu}.

\edit1{At \emph{Wind}, the \emph{Dst} index developed in three distinct steps. The first two decreases were associated with the sheath regions behind S2\&3 and S4, and the third decrease was related to E4. Both sheaths featured large amplitude fluctuations in the North--South magnetic field component ($B_{Z}$) that reached about $-10$~nT. E4 contained periods of southward field, although weak in magnitude ($B_{Z}>-5$~nT), but the solar wind speed was relatively high throughout E4, in particular at its leading edge ($V\sim700$~km$\cdot$s$^{-1}$). As a result,} both sheaths caused a moderate storm (\emph{Dst}~$\sim-60$~nT), whilst E4 drove a minor storm (\emph{Dst}~$\sim-40$~nT).

\edit1{The \emph{Dst} index estimated using STEREO-A solar wind data developed in two distinct steps. The first decrease was associated with the southward field following S1. Before S2\&3 arrived, the field magnitude decreased (and consequently also its southward component) and as a result the \emph{Dst} showed signs of recovery. The sheath behind S2\&3 featured a mostly northward field at STEREO-A and, as a consequence, the \emph{Dst} stayed at quiet-time values. The second decrease was associated with the southward field embedded in E2\&3. As this ICME was weak and slow (its speed was only about $400$~km$\cdot$s$^{-1}$), its corresponding \emph{Dst} response was quite weak.} It is noticeable that the two models gave very similar results for the sheath that followed S1 (\emph{Dst}~$\sim-50$~nT), but slightly different values of $Dst_{\text{min}}$ for E2\&3 (\emph{Dst}~$\sim-35$~nT for the Burton model versus \emph{Dst}~$\sim-45$~nT for the O'Brien model). Nevertheless, the structures associated with both CMEs at STEREO-A would likely have caused minor--to--moderate disturbances if they had impacted Earth instead.

\section{Sun--to--1~AU Connection} \label{sec:analysis}

Next, we describe the multiwavelength and multipoint analysis we perform in order to investigate how well CMEs are visible in different types of observations and from different vantage points and to connect the four CMEs from the Sun to 1~AU and assess their impact, both on the observational and on the modelling perspectives.

\subsection{Coronagraph-based Analysis\label{subsec:gcs}}

\begin{figure*}
\gridline{\fig{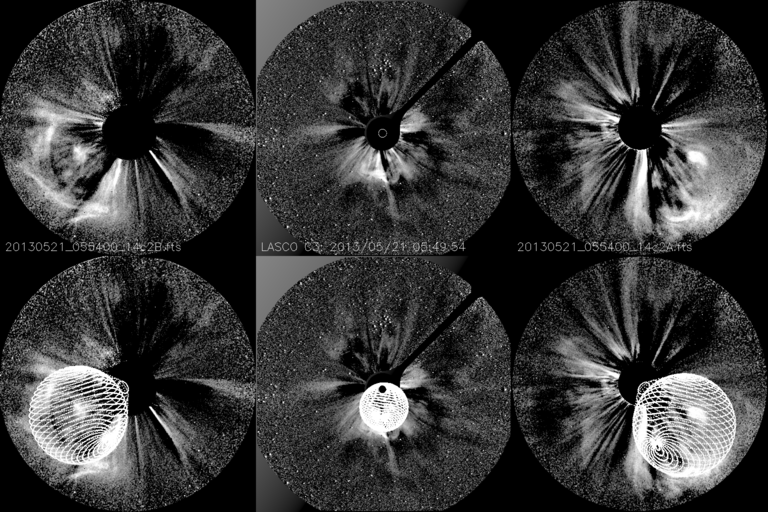}{0.45\textwidth}{(a) CME1}
         \fig{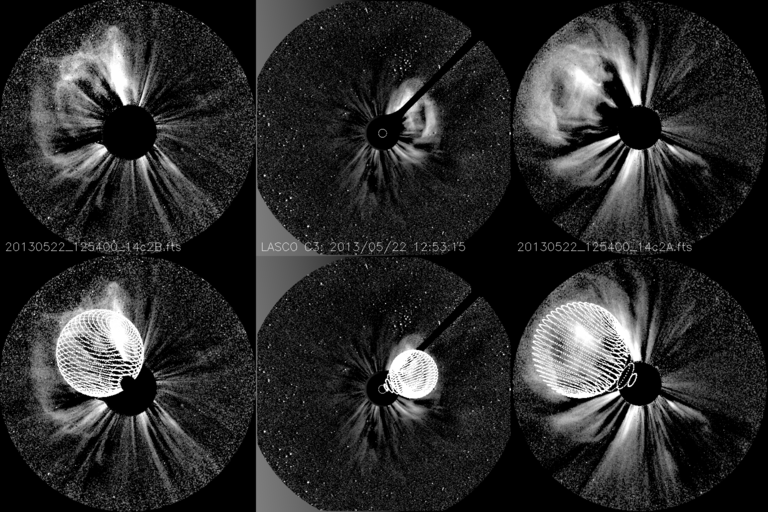}{0.45\textwidth}{(b) CME2}}
\gridline{\fig{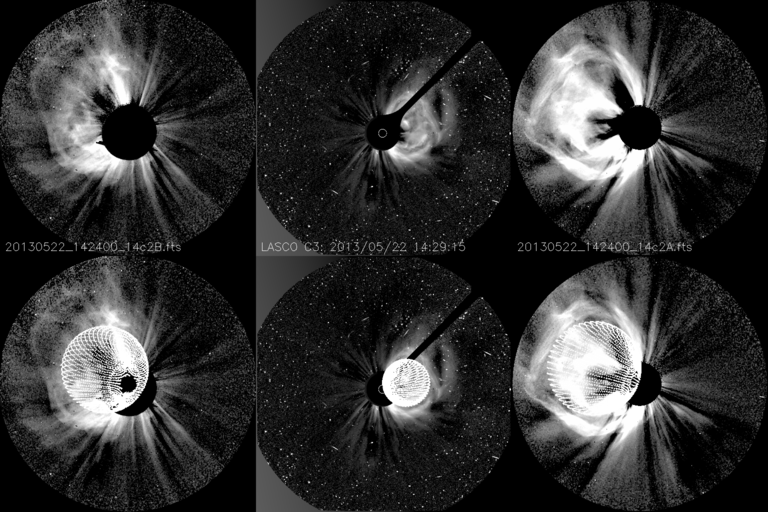}{0.45\textwidth}{(c) CME3}
         \fig{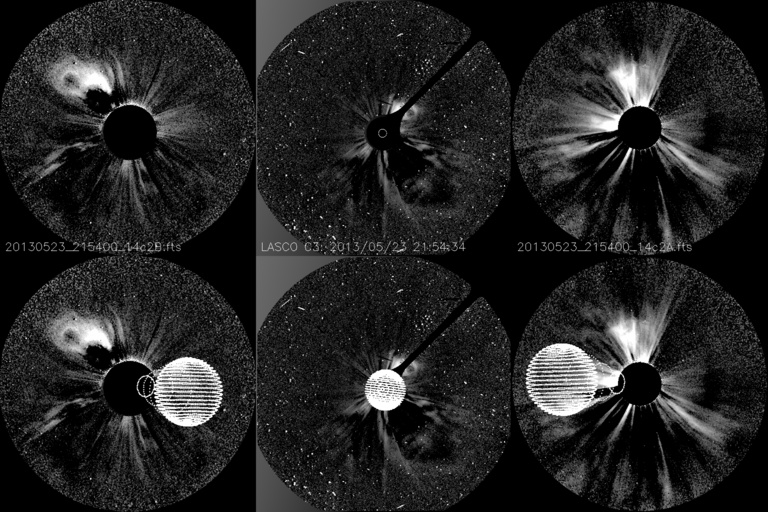}{0.45\textwidth}{(d) CME4}}
\caption{GCS cone fitting results for the four CMEs under study. The top panels in each case show multipoint coronagraph observations of the CMEs in difference images, and the lower panels show the corresponding spherical GCS fits. \label{fig:gcs}}
\end{figure*}

We perform a 3D fitting of the four CMEs in the coronagraphs from the three vantage points (SOHO, STEREO-A, and STEREO-B) using the Graduated Cylindrical Shell (GCS) model \citep{thernisien2006,thernisien2009}. The GCS fits serve as input for our propagation models (see Sections \ref{subsec:hi} and \ref{subsec:euhforia}). In order to be consistent with the assumption of a spherical shape used in the modelling, we fit all the CMEs using \edit1{the ice-cream cone model \citep{fisher1984} by setting the}\ half-angular width $\alpha=0^{\circ}$. Figure \ref{fig:gcs} shows examples of the GCS cone fittings for each case.

The parameters that we obtain as output from each GCS fitting are: latitude ($\theta$), longitude ($\phi$), height ($h$), and aspect ratio ($\kappa$, i.e.\ the ratio of the CME size at two orthogonal directions). Since we set $\alpha=0^{\circ}$ and use a cone shape, the value of the tilt angle of the CME axis ($\gamma$) is irrelevant for our fitting. We derive the CME speed ($v$) from the value of $h$ by fitting our CMEs at two separate times \edit1{and the CME half-angle ($\omega/2$) through the relation $\kappa = \sin(\omega/2)$}.

CME1, CME2, and CME3 are clearly seen from all three vantage points and, as a result, can be fitted fairly well. CME4, on the other hand, is not seen from the viewpoint of SOHO and is relatively faint in STEREO-B images. As a consequence, the GCS fit for this CME has larger uncertainties than for the other three CMEs. Furthermore, CME1, CME2, and CME3 show a sharp faint outer boundary outside the fitting of the CME bubble. These outer boundaries are often interpreted as white-light signatures of shocks that develop ahead of propagating CMEs \citep{vourlidas2013}. CME4, on the other hand, is significantly fainter in coronagraph images and, as a result, we cannot distinguish any feature that would indicate the presence of a shock.

\subsection{Radio-based Analysis} \label{subsec:shocks}

\begin{figure*}
\epsscale{1.15}
\plotone{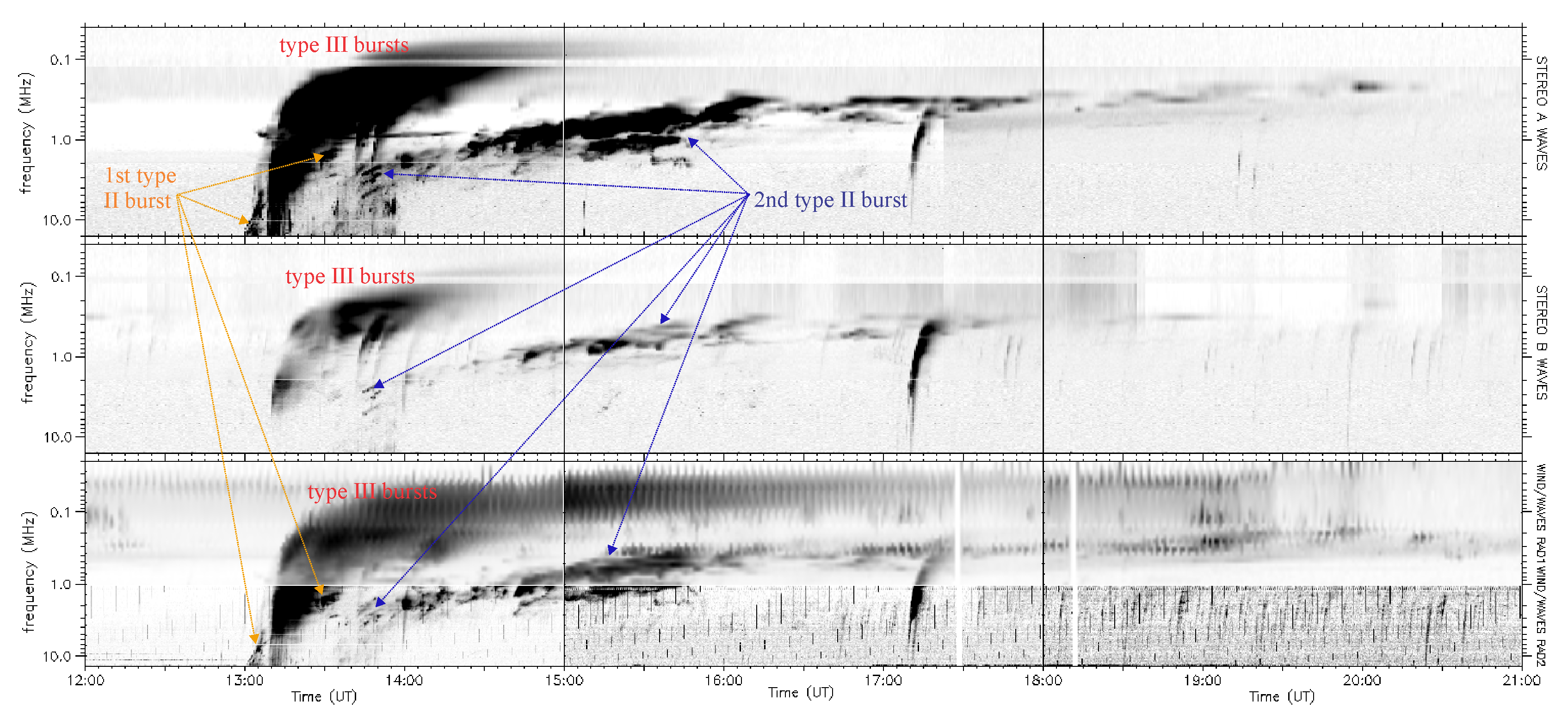}
\caption{Dynamic spectra showing the radio signatures associated with CME3 as measured by the WAVES instruments onboard STEREO-A (top), STEREO-B (middle), and \emph{Wind} (bottom). The x--axis shows the time in UT during 22~May~2013. The type II and type III bursts are indicated with arrows and labels.
\label{fig:radio}} 
\end{figure*}

Flares and CMEs are frequently associated with electromagnetic emission covering a wide spectral range. Radio emission produced by non-thermal electrons accelerated at the shock wave front, so-called type~II radio bursts \citep[e.g.,][]{wild1950,nelson1985,vrsnak2008,magdalenic2010}, are a well-known and in many aspects unique means to study the propagation of shock waves. Herein, we inspect the radio emission associated with the events under study, with the aim to obtain additional knowledge on the propagation of the CME-driven shock waves.

CME1 and CME4 were not associated with type~II radio emission. CME1 had no associated radio emission at all, neither in the metric (low corona) wavelength range, nor at the decametric--to--kilometric (interplanetary space) wavelengths, as seen from the WAVES instruments onboard \emph{Wind} and the STEREOs. CME4 was associated only with several type~III radio bursts \citep[signatures of fast electron beams propagating along open field lines; e.g.,][]{reid2014} observed by W/WAVES. CME2 also lacked radio type~II signatures in the metric range, but was associated with a short-lasting and patchy drifting emission observed by \emph{Wind}. The low frequency emission was observed from approximately 11:00 to 12:00~UT on 23~May, when CME2 was at about 10~$R_{\odot}$ away from the Sun. The observed radio emission gives indication of the presence of a shock wave associated with this rather slow CME. However, the signal-to-noise ratio of the radio emission was too weak to be analysed.

CME3, in turn, was temporally associated with a complex radio event, consisting of two adjacent type~II bursts, type~III bursts associated with the impulsive phase of the flare, and type~IV continuum (see Figure~\ref{fig:radio}). The first type~II burst was observed in the decametric--to--kilometric range from 12:55~UT until 13:35~UT, and had starting and ending frequencies of about 50~MHz in \emph{Nan{\c c}ay Decametric Array} \citep[NDA;][]{boischot1980} observations and 1.2~MHz in S/WAVES-A and W/WAVES data, respectively. S/WAVES-B observations show only the low frequency part of the emission, indicating that the shock was at the beginning of the event occulted for STEREO-B. \edit1{Since the starting frequency of this radio burst was below the observing range of the \emph{Nan{\c c}ay Radioheliograph} \citep[NRH;][]{kerdraon1997}, the comparison of the type~II source positions with EUV observations is not possible.} A second, intense type~II burst was observed in the metric range and from the Sun through the outer corona by all three WAVES instruments. It is difficult to estimate the exact starting time and frequency of this type~II burst because the metric-range counterpart occurred concurrently with the type~IV continuum. The approximate start time and frequency are estimated to be 13:11~UT and about $190$~MHz, respectively. The comparison of NRH observations with EUV data indicates that the source region of the second type~II burst was situated above the source region of CME3. By assuming that radio emission is most intense along the direction of propagation and noting the positions of the observing spacecraft depicted in Figure~\ref{fig:position}, it is possible to derive a qualitative propagation direction for the type~II radio bursts \citep[e.g.,][]{magdalenic2014}. In this regard, the first burst was mostly occulted to STEREO-B and the second one was most intense from the viewpoints of STEREO-A and \emph{Wind}, suggesting that both emissions originated and propagated roughly between Earth and STEREO-A, i.e., from the same source and in the same direction as CME3. Both type~II bursts appear slightly more intense in S/WAVES-A, suggesting a modest asymmetry in the propagation direction of the source towards STEREO-A. Different radio shock signatures may originate from different parts of a single, very extended shock wave \citep[e.g.][]{morosan2019}, or from two subsequent eruptions from the same source region happening very close in time. Since the drift rates of these two type~II bursts are very similar and we observe clear signatures of only one CME, the possibility that they are driven by the same shock wave seems plausible.

\begin{figure}
\epsscale{1.15}
\plotone{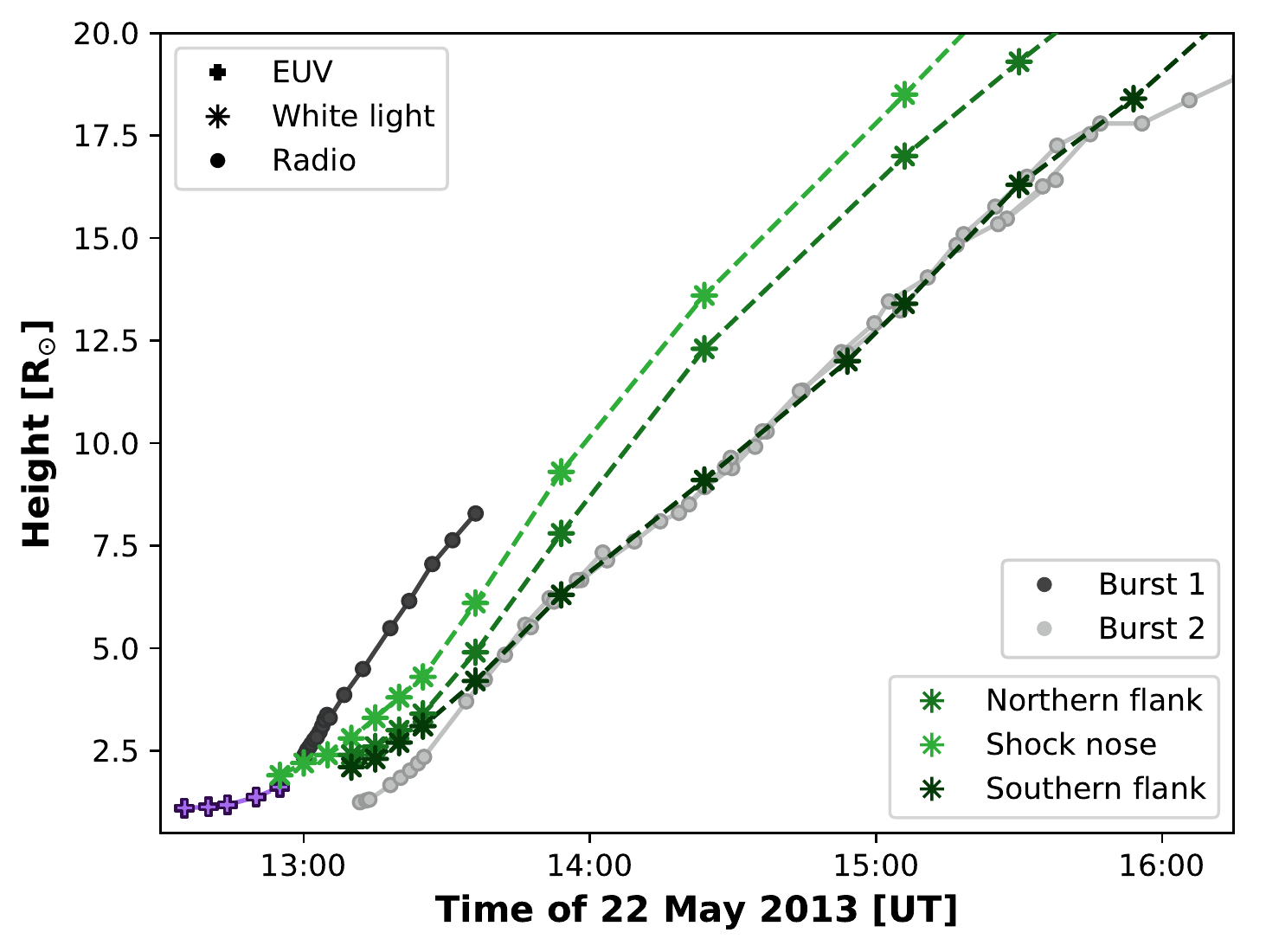}
 \caption{Kinematics of the shock associated with CME3 in \edit1{EUV}, white light, and radio. EUV measurements (plus symbols) are taken with PROBA2/SWAP and STEREO/SECCHI/EUVI-A, white-light measurements (asterisks) are taken with SOHO/LASCO and STEREO/SECCHI/COR-A, and radio measurements (circles) are taken from NDA in the metric range and from the WAVES instruments onboard \emph{Wind} and both STEREO spacecraft in the decametric--to--kilometric range. \label{fig:cme3shock}}
\end{figure}

\edit1{To attempt to define the exact sources of the radio emission associated with CME~3, we study the height--time profiles of the type~II bursts and compare them with the kinematics of the expanding CME observed in EUV and white light. Regarding EUV observations, we track the erupting CME loop off limb. Regarding white-light observations, we track the position of the shock nose and its northern and southern flanks. 
We obtain the kinematics of the type~II radio bursts from the frequency drift rate using the two-fold Saito \citep{saito1970} coronal density model for ground-based observations and the hybrid model developed by \citet{vrsnak2004} for space-based observations. The hybrid model provides a smooth transition from the active region corona to interplanetary space, and is therefore applicable for radio observations in the decameter to kilometer range. Although the type~II bursts show a complex morphology, it is possible to determine whether the radio emission was at the fundamental or the harmonic of the plasma frequency. The harmonic emission lanes are converted to fundamental emission (through division by 2) before the conversion into heights.}

\edit1{The resulting kinematic evolution is shown in Figure~\ref{fig:cme3shock}, from which it is evident that the second type~II radio burst likely originated from close to the southern flank of the shock driven by CME3, in agreement with previous studies \citep[e.g.,][]{reiner1998,magdalenic2014,martinezoliveros2015,krupar2016}. The first type~II burst appears to correlate well with the kinematic curve of the CME3-driven shock nose in its early stage, but later deviates from the shock nose kinematics. Keeping in mind the uncertainties associated to the selection of the density models in radio studies, together with the projection effects and uncertainties arising from tracking shock waves in white-light, we conclude that the first burst likely originated from close to the nose of the shock driven by CME3. From the time--height points shown in Figure~\ref{fig:cme3shock}, we derive the speeds of the observed type~II during their initial propagation (until $\sim10R_{\odot}$) to be $\sim1900$~km$\cdot$s$^{-1}$ for the first one and $\sim1300$~km$\cdot$s$^{-1}$ for the second one. The derived speeds match well with the kinematics of the CME3 ejecta in the corona (we obtained a speed of $\sim1500$~km$\cdot$s$^{-1}$ from the GCS reconstruction presented in Section~\ref{subsec:gcs}).}

\subsection{HI-based Analysis} \label{subsec:hi}

\begin{figure*}
\epsscale{1.1}
\plotone{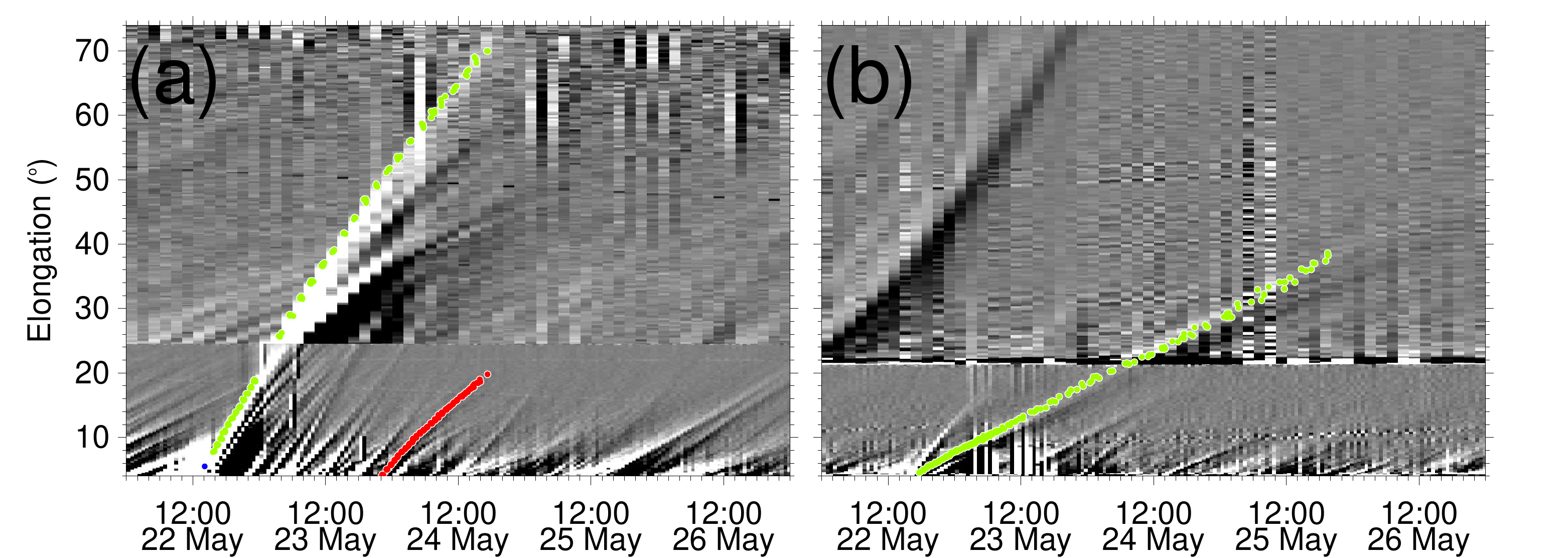}
\caption{Time--elongation data for the period 22~May to 26~May~2013. Panels (a) and (b) show HI running-difference time--elongation maps for STEREO-A and -B, respectively, constructed along position angles of $85^{\circ}$ and $295^{\circ}$. The overplotted circles show the manually tracked time--elongation profiles for CME2 (blue), CME3 (green), and CME4 (red). Earth was located at about $20^{\circ}$ in both fields of view. \label{fig:jmap}}
\end{figure*}

Next, we follow the CMEs in the HI cameras onboard both STEREO spacecraft, which each have a combined elongation coverage between $4^\circ$ and $87.7^\circ$ between the Sun and Earth, centred on the ecliptic. As discussed in Section~\ref{subsec:remotesensing}, CME1 was directed roughly between STEREO-A and STEREO-B, away from Earth, thus it did not pass through the HI fields of view. We did, however, observe the remaining CMEs. 

Because CME2 was only visible in two images from HI1-A before it was subsumed by CME3, we take the plane of sky location of the front of CME2 to calculate its height. CMEs2\&3 were subsequently visible as a single CME in the observations from HI on STEREO-A and -B (see Section~\ref{subsec:remotesensing}) and so we treat them as a single entity in our further analysis. We track the elongation of the shock front of this merged CME in the ecliptic as a function of time through the HI fields of view in each spacecraft. This tracking is carried out using time--elongation maps \citep[e.g.,][]{sheeley2008,davies2009} of running difference images from HI at a position angle in the ecliptic plane. In these time--elongation maps, a propagating density structure such as a CME (and in particular the dense sheath region ahead of it) appears as a bright front followed by a dark front. The elongation of the CME front, as a function of time, is tracked by manually selecting points along the front. Due to the presence of the galactic plane in the HI1-B field of view (see Figure~\ref{fig:hi}), these images are more difficult to interpret and as a result CME4 was visible in HI on STEREO-A only. Again, we track the elongation of this CME front in the ecliptic as a function of time, albeit with data just from HI on STEREO-A. In both cases, the features are tracked to the furthest elongation possible, until they become too faint to distinguish from the background: $69^\circ$ (STEREO-A) and $38^\circ$ (STEREO-B) for CMEs2\&3 and $19^\circ$ for CME4. The structures that we track in the HI cameras from both STEREO spacecraft are shown in Figure~\ref{fig:jmap}. Earth was located at $22^{\circ}$ (STEREO-A) and $19^{\circ}$ (STEREO-B) elongation during the observation time.

Since the features tracked in the HI cameras are highly sensitive to projection effects, we apply to the time--elongation points a self-similar expansion model that allows us to resolve the tracked front in 2D. To the data for CMEs2\&3, tracked from both STEREO spacecraft, we apply the \emph{Stereoscopic Self-Similar Expansion} \citep[SSSE;][]{davies2013} model, which assumes that the CME front is represented by a circle with a constant half-width. We derive the position of the CME apex as a function of time and fit a second order polynomial, which allows us to interpolate the CME propagation and estimate an exact time of impact at Earth. For CME4, which was visible in HI1-A only, we apply the single-spacecraft version of the SSSE model, i.e.\ the \emph{Self-Similar Expansion} \citep[SSE;][]{davies2012} model. This model also represents the CME front as a circle with a constant half-width and makes the further assumptions that it propagates radially and with a constant speed. In the use of both fitting models, we apply the correction derived by \citet{mostl2013}, whereby the CME flank is expanding at a slower speed along the Sun--Earth line than it is at the apex. In order to account for the fact that a CME-driven shock is expected to be somewhat larger than the CME itself \citep[e.g.,][]{gopalswamy2010a,good2016}, we apply the (S)SSE models to a range of half-widths between the CME half-width as derived from the GCS fitting ($(\omega/2)_{\text{GCS}}$, $64^\circ$ for CMEs2\&3 and $19^\circ$ for CME4, see Section~\ref{subsec:gcs}) and  $(\omega/2)_{\text{GCS}}+15^{\circ}$, using $5^{\circ}$ increments. The resulting impact times and speeds for different half-widths are shown in Table~\ref{tab:himodel}.

\begin{table}
\caption{Arrival speeds and times of the CMEs tracked in HI time--elongation data and fitted through the SSSE (CMEs2\&3) and SSE (CME4) models, using different half-widths. The columns show, from left to right: CME number, half-width, speed of the front apex at the time of impact at Earth, speed of the front along the Sun--Earth line at the time of impact, arrival time at Earth in UT and in the format DD/HH:MM, and angle between the CME apex and the Sun--Earth line (the plus sign is defined towards the West).}
\label{tab:himodel}
\centering
\begin{tabular}{c c c c c c}
\hline\hline
CME & $\omega/2$ & $V_{\text{Apex}}$ & $V_{\text{Earth}}$ & $t_{\text{Earth}}$ & $\delta$ \\
\hline
2\&3 & $64^{\circ}$ & 1095 km/s & 792 km/s & 24/13:39 & $-38^{\circ}$\\
2\&3 & $69^{\circ}$ & 1120 km/s & 810 km/s & 24/14:49 & $-39^{\circ}$\\
2\&3 & $74^{\circ}$ & 1141 km/s & 825 km/s & 24/15:48 & $-39^{\circ}$\\
2\&3 & $79^{\circ}$ & 1138 km/s & 822 km/s & 24/16:18 & $-40^{\circ}$\\
\hline
4 & $19^{\circ}$ & 963 km/s & 952 km/s & 25/13:58 & $-5^{\circ}$\\
4 & $24^{\circ}$ & 990 km/s & 989 km/s & 25/12:23 & $-1^{\circ}$\\
4 & $29^{\circ}$ & 1016 km/s & 1015 km/s & 25/11:18 & $+2^{\circ}$\\
4 & $34^{\circ}$ & 1043 km/s & 1035 km/s & 25/10:31 & $+5^{\circ}$\\
\hline\hline
\end{tabular}
\end{table}

By checking the angle $\delta$ shown in Table~\ref{tab:himodel} between the shock apex resulting from the fittings and the Sun--Earth line, we note that all the reported values result in an eastern flank encounter for CMEs2\&3 and a frontal encounter for CME4, as expected from the observational analysis reported in Section~\ref{sec:overview}. CMEs2\&3, in fact, are associated with values of $\delta$ around $-39^{\circ}$, whilst for CME4 all the fittings result in an impact at Earth within $\pm5^{\circ}$ from the shock apex. Furthermore, in both cases, the fittings that assume a half-width as derived from the GCS reconstructions yield an impact time at Earth that lies about $4$~hours from the actual shock arrival ($\sim-4$~hours for CMEs2\&3 and $\sim+4.5$~hours for CME4), with both fittings approaching the measured in-situ arrival time with increasing half-width. The half-width of $(\omega/2)_{\text{GCS}}+15^{\circ}$ gives the closest impact time at Earth when compared to in-situ data in both cases ($\sim-1$~hour for CMEs2\&3 and $\sim+1$~hour for CME4), suggesting that the larger extent of a CME-driven shock should be taken into account when tracking and reconstructing features observed in the HI cameras. However, the SSSE model relies on several assumptions, which oversimplify the true CME physics. Primarily, the model assumes the cross-section of the CME front to be a rigid, expanding circle. It is not possible for a circular front constrained by the observed time--elongation data for CMEs2\&3 to ever reach STEREO-A. However, a flatter shock front that is constrained by the same data would have a wider longitudinal range and may indeed be capable of reaching STEREO-A. Furthermore, the circular fronts assumed in the (S)SSE model result in arrival speeds at 1~AU that are much larger than the observed ones at \emph{Wind} ($\Delta V\sim200$~km$\cdot$s$^{-1}$ for CMEs2\&3 and $\Delta V\sim400$~km$\cdot$s$^{-1}$ for CME4).

\subsection{EUHFORIA Modelling} \label{subsec:euhforia}

In order to assess the validity of our estimated in-situ impacts at both \textit{Wind} and STEREO-A, we run a simulation with the \emph{EUropean Heliospheric FORecasting Information Asset} \citep[EUHFORIA;][]{pomoell2018} model to compare to our observation-based analysis. The EUHFORIA model is composed of a simple semi-empirical Wang--Sheeley--Arge (WSA)-like coronal model \citep{arge2004} and a 3D time-dependent magnetohydrodynamics (MHD) heliospheric model that allows to model propagating CMEs on a steady background solar wind in the inner heliosphere from $0.1$~AU onwards. We here put the outer boundary of the computational domain at $2.0$~AU. In this work we employ a cone CME model in which CMEs are described as dense, spherical blobs of plasma injected in the heliosphere without any internal magnetic field structure, i.e.\ their magnetic field is just that of the background solar wind \citep{odstrcil2004,scolini2018b}. Due to the simplified treatment of the CME structure, cone models are not suitable to study the magnetic field structure associated with ICMEs. However, they have been successfully applied to study the global evolution of CMEs and their fronts in the heliosphere, and to predict CME arrival times at different locations.

As input for the semi-empirical coronal model we use the synoptic standard magnetogram generated by the \emph{Global Oscillation Network Group} (GONG) on 21~May~2013 at 08:14~UT. This allows us to model the background solar wind conditions before the eruption of CME1.

We simulate the four CMEs under study by inserting them at the heliospheric inner boundary, set at $0.1$~AU (corresponding to 21.5~$R_{\odot}$). The input parameters for each CME were derived from the GCS reconstructions at the latest time when each CME was visible in the coronagraphs, e.g.\ as close as possible to the inner boundary of EUHFORIA (see Table~\ref{tab:euhforia_cme_input}). 
\begin{table}
\caption{CME input parameters used in EUHFORIA, as derived from the GCS modelling.
$\theta$ and $\phi$ are in Stonyhurst coordinates.}
\label{tab:euhforia_cme_input}
\centering
\begin{tabular}{l r r r r}
\hline\hline
& CME1 & CME2 & CME3 & CME4 \\
\hline
Date             & 2013-05-21  & 2013-05-22 &  2013-05-22 & 2013-05-23 \\
Time             & 07:31~UT & 15:34~UT &  15:40~UT & 22:25~UT \\
$v$             & 735~km/s & 541~km/s & 1507~km/s & 1430~km/s \\
$\theta$        & -25$^\circ$ & 30$^\circ$  &20$^\circ$  & -2$^\circ$  \\
$\phi$          & 174$^\circ$ & 68$^\circ$ & 70$^\circ$  & 3$^\circ$   \\
$\omega/2$      & \edit1{38$^\circ$} & \edit1{41$^\circ$}  & \edit1{64$^\circ$} & 19$^\circ$  \\
\hline\hline
\end{tabular}
\end{table}

\begin{figure}
\centering
{\includegraphics[width=.99\linewidth,trim={10mm 144mm 0mm 00mm},clip]{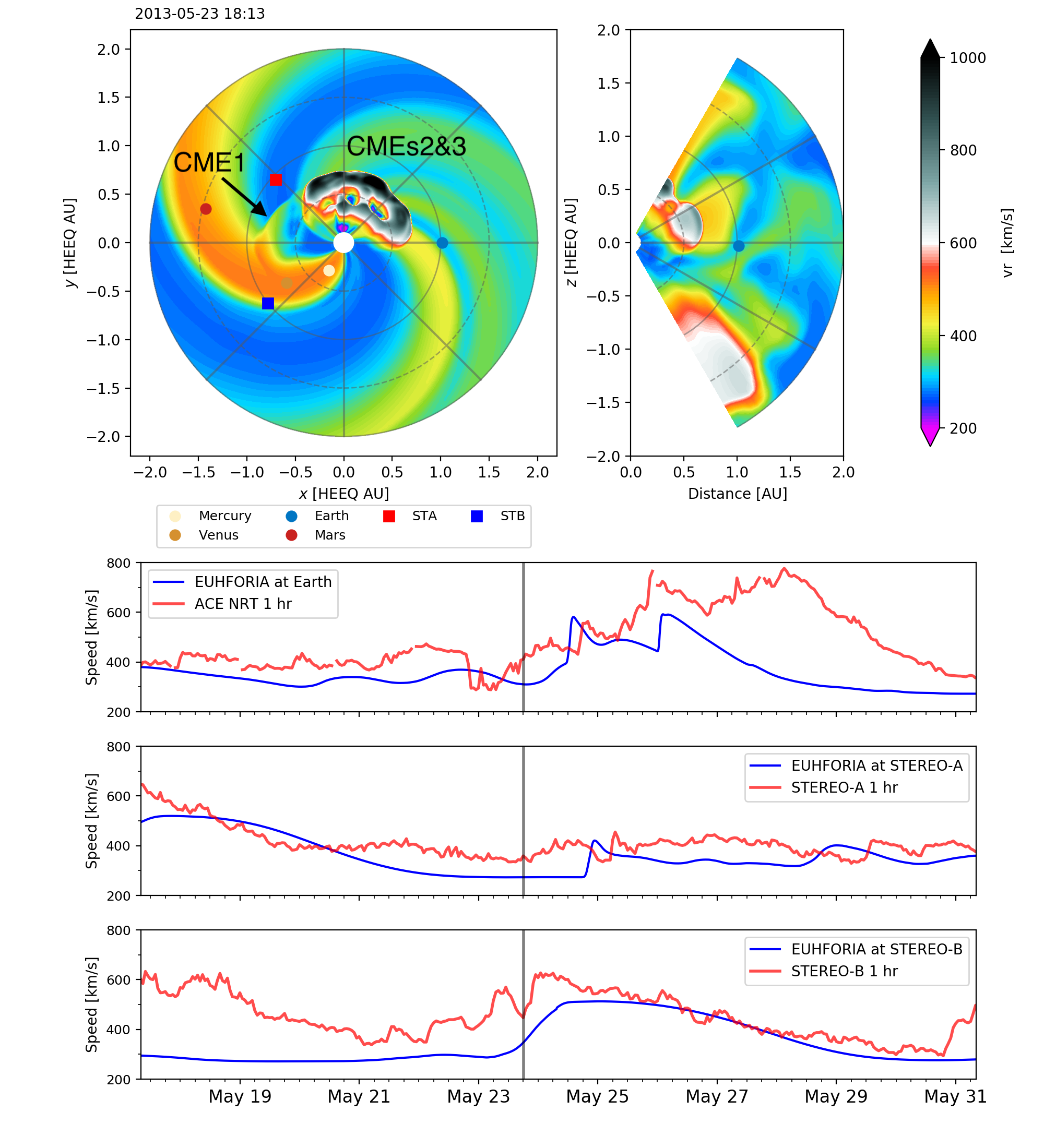} \\ 
\includegraphics[width=.99\linewidth,trim={10mm 144mm 0mm 00mm},clip]{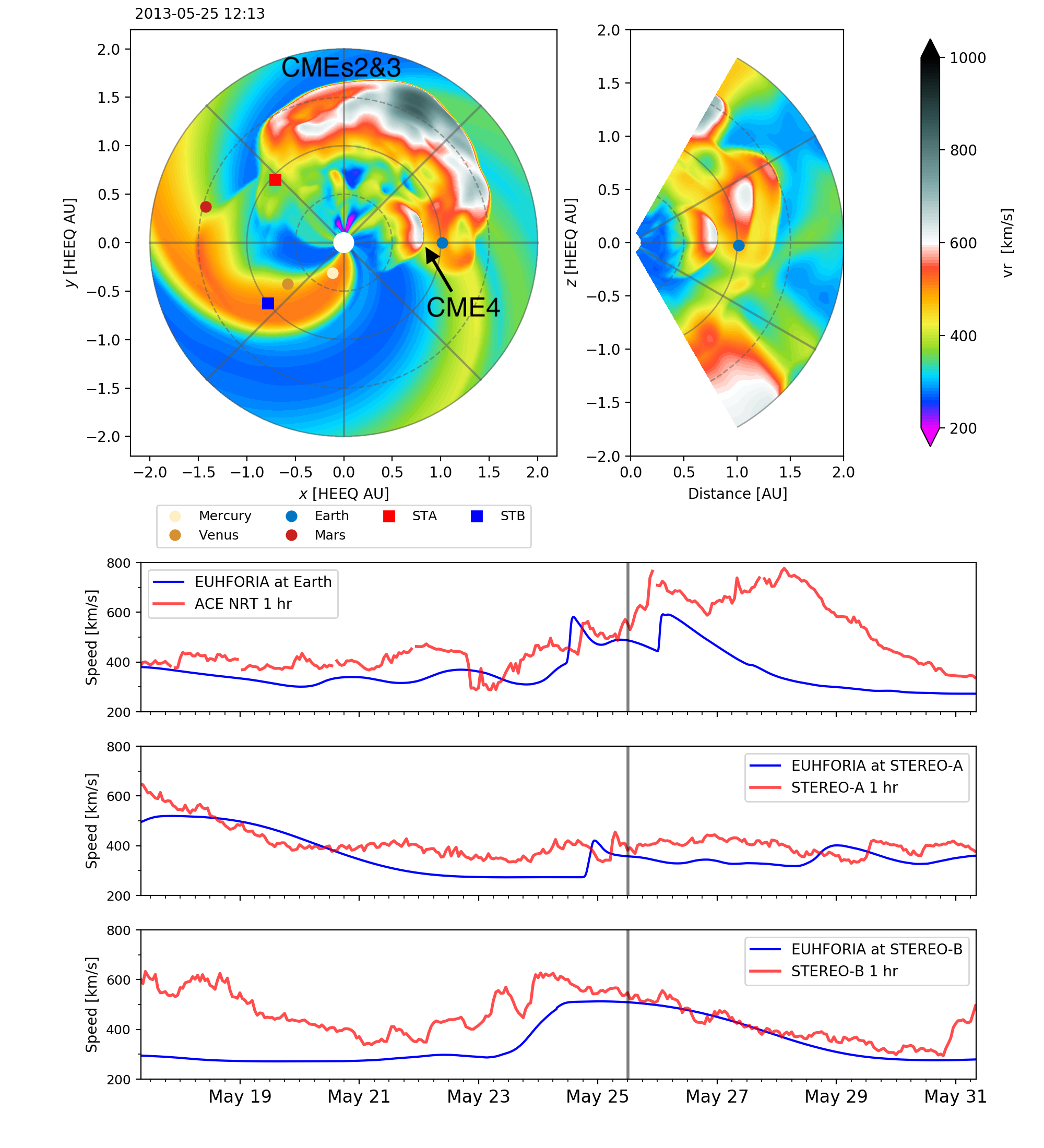} \\
\includegraphics[width=.99\linewidth,trim={10mm 0mm 0mm 135mm},clip]{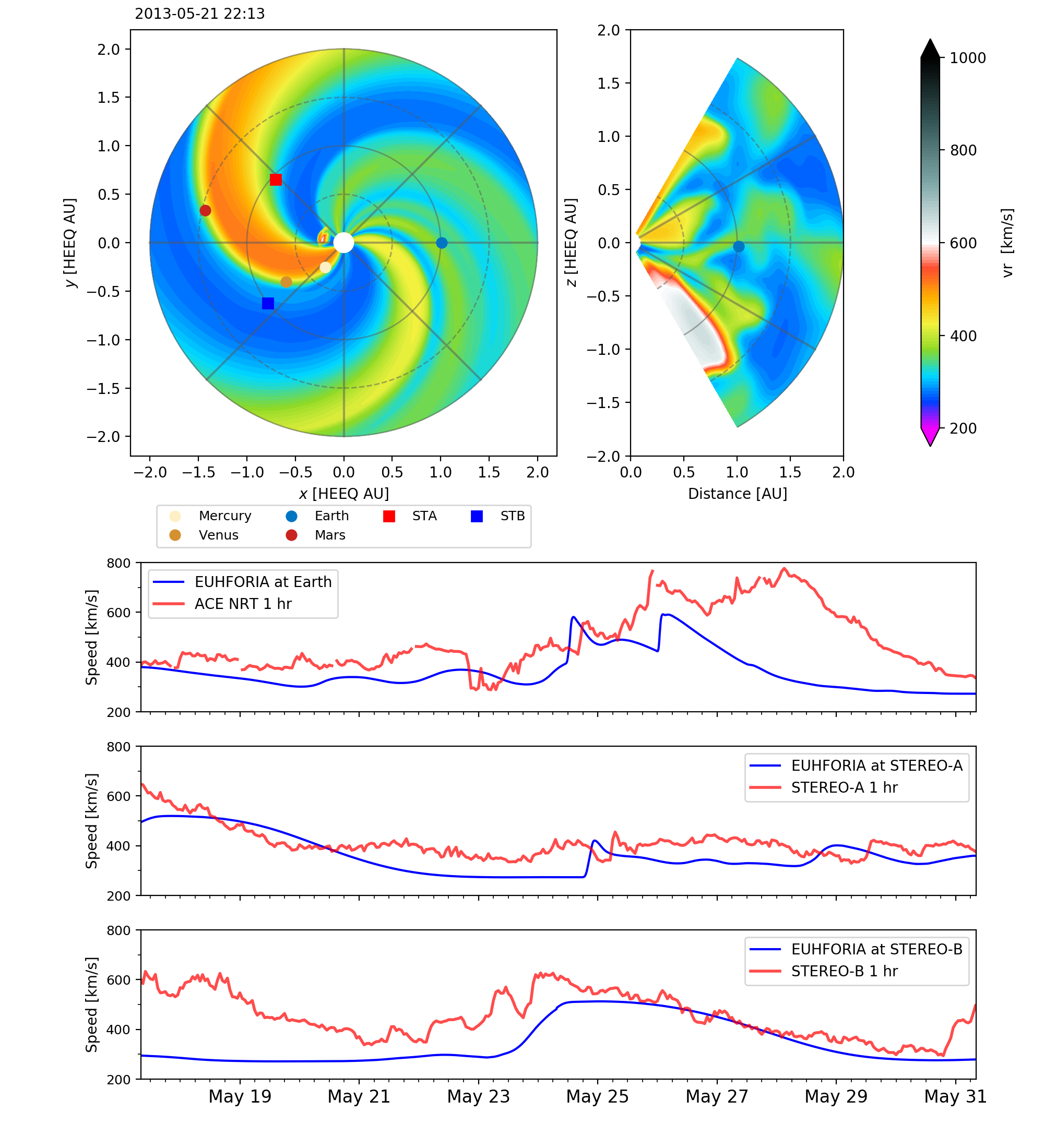}} 
\caption{Top: Snapshots of the EUHFORIA simulation at 18:13~UT on 23~May~2013 and at 12:13~UT on 25~May~2013, in the heliographic equatorial plane and in the meridional plane that includes Earth. Bottom: comparison of the EUHFORIA time series (blue) with in-situ measurements from ACE, STEREO-A, and STEREO-B (red), during the whole temporal computational domain.\\
(An animation of this figure is available.) \label{fig:euhforia}} 
\end{figure}

Figure~\ref{fig:euhforia} shows two snapshots of the EUHFORIA simulation for the speed in the heliographic equatorial plane and meridional plane containing Earth, and a comparison of the EUHFORIA prediction at Earth, STEREO-A, and STEREO-B with in-situ measurements from the \textit{Advanced Composition Explorer} \citep[ACE;][]{stone1998} spacecraft and from the two STEREO spacecraft. As is visible in the top panels, CME1 is modelled as back-sided with respect to Earth, propagating mainly between STEREO-A and STEREO-B.
CME2 and CME3 merge during their insertion at the inner boundary of the model, and are therefore observed as a single structure propagating between Earth and STEREO-A. Finally, CME4 is approximately Earth-directed, suggesting an almost central hit at Earth.

As can be seen in the top panel of Figure~\ref{fig:euhforia}, the model predicts that the longitudinal extent of CME1 encompasses the position of STEREO-A, suggesting an impact (albeit weak) of CME1 on STEREO-A at its eastern flank. 
However, in the simulation the merged CMEs2\&3 propagate at a faster speed, catching up and merging with CME1 shortly before the impact at STEREO-A (second panel) at around 18:00~UT on 24 May 2013. 
For this reason, the speed time series at STEREO-A shows a single peak, corresponding to the arrival of CME1 together with CMEs2\&3, instead of two individual peaks. A comparison of the simulation results with in-situ observations at STEREO-A indicates that the arrival of CMEs2\&3 in the simulation occurs about 12 hours earlier than observed. As CME1 is modelled to arrive together with CMEs2\&3 at STEREO-A, its arrival is about 12 hours late in the model.
%
At Earth, the EUHFORIA time series shows the presence of two peaks. The first peak corresponds to the arrival of the merged CMEs2\&3, occurring around 12:00~UT on 24 May, i.e.\ less than 6 hours earlier than observed from in-situ measurements. The second peak corresponds to the arrival of CME4, expected around 00:00~UT on 26 May. In this case, the EUHFORIA prediction is about 14~hours late when compared to in-situ data.
%
Finally, none of the modelled CMEs is predicted to arrive at STEREO-B, as confirmed by the corresponding speed time series. The increase in solar wind speed (from $V\sim300$~km$\cdot$s$^{-1}$ to $V\sim500$~km$\cdot$s$^{-1}$) is to attribute to a fast wind stream that, according to our simulation, impacted STEREO-B at around the time shown in the top panel of Figure~\ref{fig:euhforia}.

In summary, EUHFORIA predicts CME1 to arrive at STEREO-A, CMEs2\&3 to arrive at both STEREO-A and Earth, and CME4 to arrive at Earth. Moreover, CME1 is not seen to arrive at STEREO-B. The CME-driven shock speeds at 1~AU are consistent with in-situ measurements, with errors ranging between $\sim5$ and $\sim30$~km$\cdot$s$^{-1}$ only. Despite some discrepancies between predicted and observed arrival times of the CMEs at various spacecraft, compatible with typical uncertainties of similar models (\citealp{riley2018,wold2018}; possibly due to inaccuracies in the modelling of the solar wind background and in the CME speed at 21.5~$R_{\odot}$ derived from GCS fitting), the prediction of the impacts in the EUHFORIA simulation is consistent with the analysis presented in Section~\ref{subsec:remotesensing} that was based on remote-sensing observations of the events under study, as well as with the in-situ observations discussed in Section~\ref{subsec:insitu}.

\section{Discussion and Conclusions} \label{sec:conclusions}

In this work, we have performed a Sun--to--1~AU analysis of four CMEs that erupted during 21--23~May~2013. These eruptions took place during a period in which the Sun was very active (at about six CMEs per day), but we uniquely matched these four events with their in-situ signatures through detailed examination of observational data supported by modelling. All the CMEs under study could be considered ``problematic'' from a space weather forecasting perspective because they either erupted from the solar limb with respect to their target location (CME1, CME2, and CME3) or they did not have clear signatures in coronagraph images (CME4). Nevertheless, all the CMEs that arrived at Earth caused a moderate geomagnetic disturbance, and we estimated using \emph{Dst} prediction models that the CMEs observed by STEREO-A would likely have also caused moderate activity had they impacted Earth instead.

The moderate storm activity revealed by the \emph{Dst} index related to the analysed events was associated at both \emph{Wind} and STEREO-A to turbulent sheath fields behind interplanetary shocks and relatively weak ICME ejecta. The ejecta observed at \emph{Wind} (E4 in Figure~\ref{fig:insitu}(a)) was likely weak due to its being associated to a narrow and faint solar counterpart, whilst the ejecta detected at STEREO-A (E2\&3 in Figure~\ref{fig:insitu}(b)) corresponded to the glancing encounter with a large, merged CME that was launched roughly between Earth and STEREO-A.

Three of the four CMEs under study (CME1, CME2, and CME3) erupted from the solar limb, with their source regions located at $> \pm 65^{\circ}$ longitude from Earth's and/or STEREO-A's viewpoints. The fact that they caused moderate space weather disturbances at 1~AU agrees with a number of previous studies \citep[e.g.,][]{rodriguez2009,gopalswamy2010b,cid2012}. These disturbances were caused primarily through the CMEs' sheath regions, although we note that in the case of CMEs2\&3 at STEREO-A a minor disturbance was triggered by the edge-encountered ejecta instead. Note that CME2 and CME3 erupted from the eastern limb as seen from STEREO-A, which could partly explain their weak geomagnetic response. West-limb CMEs tend to be more geoeffective due to eastward deflection driven by the spiral nature of solar wind structures \citep{gosling1987,wang2004,wang2014}. However, \citet{gopalswamy2010b} found that CMEs originating close to the eastern limb can cause geomagnetic storms under extreme conditions (i.e., when speeds exceed $2000$~km$\cdot$s$^{-1}$). The largest and fastest East-limb CME in our study (CME3) had a speed $\sim1500$~km$\cdot$s$^{-1}$ (from the GCS reconstruction) close to the Sun, i.e. below the \citet{gopalswamy2010b} limit. To summarise, the results presented in this work suggest that limb CMEs can be geoeffective as a result of deflections in the corona (as in the case of CME1) or because of large longitudinal extents (as for CMEs2\&3).

\begin{table*}
\caption{Summary of the remote-sensing and in-situ observations and reconstructions, together with heliospheric modelling results, related to the four CMEs under study. The dates are shown in the format DD/HH:MM.}
\label{tab:summary}
\centering
\begin{tabular}{|l|c|c|c|c|}
\hline\hline
 & CME1 & CME2 & CME3 & CME4 \\
\hline\hline
Eruption time & 21/01:00 & 22/07:00 & 22/12:30 & 23/19:30 \\
Source from Earth & Back-sided & NW limb & NW limb & Centre \\
Source from STEREO-A & SW limb & NE limb & NE limb & Back-sided \\
\hline
Direction in coronagraph & STEREO-A/STEREO-B & Earth/STEREO-A & Earth/STEREO-A & Earth \\
Speed in coronagraph & 735 km/s & 541 km/s & 1507 km/s & 1430 km/s \\
Shock in coronagraph & Yes & Yes & Yes & No \\
\hline
Type II burst & No & Yes (too faint) & Yes (2 bursts) & No \\
Direction of type II & --- & --- & Earth/STEREO-A & --- \\
Speed of type II & --- & --- & 1939 \& 1268 km/s & --- \\
\hline
Observed in HI & No & A only & A \& B & A only \\
(S)SSE hit at 1 AU & No & \multicolumn{2}{c|}{Earth (E flank)} & Earth (Nose) \\
Arrival at Earth & --- & \multicolumn{2}{c|}{24/16:18 (822 km/s)} & 25/10:31 (1035 km/s) \\ 
\hline
EUHFORIA hit at 1 AU & STEREO-A (E flank) & \multicolumn{2}{c|}{Earth (E flank) \& STEREO-A (W flank)} & Earth (Nose) \\
Arrival at Earth & --- & \multicolumn{2}{c|}{24/12:44 (600 km/s)} & 26/00:55 (590 km/s) \\
Arrival at STEREO-A & 24/20:38 (430 km/s) & \multicolumn{2}{c|}{24/20:38 (430 km/s)} & --- \\
\hline
Observations at \emph{Wind} & No & \multicolumn{2}{c|}{Shock} & Shock+Ejecta \\
Shock arrival & --- & \multicolumn{2}{c|}{24/17:26 (590 km/s)} & 25/09:22 (580 km/s) \\
Ejecta arrival & --- & \multicolumn{2}{c|}{---} & 25/23:58 (705 km/s) \\
\hline
Observations at STEREO-A & Shock & \multicolumn{2}{c|}{Shock+Ejecta} & No \\
Shock arrival & 24/06:52 (425 km/s) & \multicolumn{2}{c|}{25/06:05 (460 km/s)} & --- \\
Ejecta arrival & --- & \multicolumn{2}{c|}{25/17:06 (410 km/s)} & --- \\
\hline\hline
\end{tabular}
\end{table*}

CME4, on the other hand, erupted from close to the disc centre as seen from Earth, but we defined it as a ``problematic'' CME nevertheless. Although its on-disc eruption signatures were clear, CME4 was not visible in LASCO imagery and its morphology in STEREO-B would have been classified as a ``jet'' according to the definition of \citet{vourlidas2013}, where a jet is defined as a narrow CME lacking a sharp front, a detailed sub-structure, or circular morphology. Only in the STEREO-A field of view did this CME have a three-part structure \citep{illing1985}, albeit relatively faint. We could reconstruct an approximate propagation speed and direction for CME4 only through the two STEREO viewpoints; it would have been impossible to obtain such information if we were relying on SOHO observations only. This highlights the importance of multipoint observations for understanding and forecasting events that lack clear signatures in remote-sensing data. This also applies to so-called ``stealth CMEs'' \citep[e.g.,][]{robbrecht2009,nitta2017}, which are usually detected in coronagraph imagery but lack unambiguous low-coronal counterparts. It has been shown that some CMEs that are not observed on disc can nevertheless be detected off limb from a second viewpoint \citep[e.g.,][]{robbrecht2009,vourlidas2011}, supporting the conjecture that stealth CMEs are not fundamentally different from ``standard'' CMEs and that their stealthiness is due to observational limitations \citep[e.g.,][]{howard2013,lynch2016b}. Likewise, CMEs that are not visible in coronagraph imagery from one viewpoint could be detected from a second one, as demonstrated in this study for CME4. Now considering the STEREO-A viewpoint only, the fact that CME4 was not visible in COR1 and then became increasingly clear through COR2 until HI1 agrees with the conclusions drawn by \citet{simnett2005} and \citet{howard2008b}, i.e.\ that some CMEs gain excess mass compared to the ambient solar wind progressively during their propagation. Mass accretion in the heliosphere has indeed been demonstrated in several studies \citep[e.g.,][]{lugaz2005,deforest2013}, although the CME white-light enhancements observed in remote measurements may also be influenced by Thomson scattering effects.

The key results of the multispacecraft and multiwavelength analysis that we have performed in order to link on-disc and in-situ observations, together with heliospheric modelling results, are summarised in Table~\ref{tab:summary}. First, we note that even though three CMEs (CME1, CME2, and CME3) showed shock signatures in white-light observations, only one of them (CME3) was associated with clear type~II radio bursts. This is consistent with the results presented by \citet{reiner2007}, who concluded that CMEs with initial speeds exceeding $1000$~km$\cdot$s$^{-1}$ are most likely to generate type~II emission. In the case of CME3, thanks to the favourable positions of three well-separated spacecraft carrying radio antennas (longitudinal separations are provided in the caption of Figure~\ref{fig:position}), we could infer the propagation direction of the associated type~II emission, which appeared to be between Earth and STEREO-A. Furthermore, the slight asymmetry in the propagation direction towards STEREO-A may explain why an ICME ejecta was observed in situ at STEREO-A, whilst only the corresponding CME-driven shock was detected at \emph{Wind}. Hence, similarly to multipoint EUV and white-light observations, multipoint radio observations are also highly important for understanding eruptive events as they allow us to estimate the propagation direction of the associated CME and associated shock. Doing so, we have clearly demonstrated that the two subsequent type~II bursts observed from the metric to the decametric ranges can be signatures of the same CME-driven shock wave. This finding also shows, in one single event, that both shock-nose and shock-flank regions can be sources of radio emission, indicating electron acceleration at multiple locations of the CME shock, in agreement with previous studies \citep{morosan2019}.

HI-based observations and reconstructions, on the other hand, were not favoured by the spacecraft configuration during the period under analysis (Figure~\ref{fig:position}). STEREO-B images were contaminated by the presence of the galactic plane, and the longitudinal distance between Earth and STEREO-A ($137^{\circ}$) was too large to result in an impact of the merged CMEs2\&3 at STEREO-A under the assumption of a circular front in the SSSE model. CME and CME-driven shock fronts tend to flatten during interplanetary propagation as a result of solar wind drag \citep[e.g.,][]{vrsnak2013}, hence HI-based reconstructions that employ an elliptical front \citep[e.g.,][]{rollett2016} would be more appropriate. The front of CME4 was likely less flattened because of solar wind preconditioning \citep[e.g.,][]{liu2014,temmer2015} due to CMEs2\&3, but its propagation direction was subject to larger uncertainties because of the single viewpoint (from STEREO-A). Nevertheless, albeit large errors in the impact speeds likely resulting from the circular-front assumption, the arrival times at Earth from (S)SSE fittings were fairly consistent with in-situ observations, with $\Delta t\sim4$~hours in both cases when a half-angle that equals the CME half-angle (from GCS reconstructions) was used. The discrepancies in arrival times were reduced to $\Delta t\sim1$~hour when $15^{\circ}$ were added to the CME half-angle, in order to account for the larger extent of the CME-driven shock. Furthermore, we were able to obtain through (S)SSE reconstructions the correct impact locations at Earth with respect to the shock nose, i.e., eastern flank for CMEs2\&3 and close to the nose for CME4. Thus, the results presented here underscore the necessity and utility of heliospheric imaging of the Sun--Earth line for terrestrial space weather forecasting.

The EUHFORIA simulation results confirmed that each of the CMEs impacted their expected spacecraft at 1~AU, albeit with some discrepancies in the arrival times (ranging between $5$ and $14$~hours) relative to the in-situ measurements. The shock arrival speeds, in turn, were significantly more consistent with in-situ measurements (the maximum error was $\sim30$~km$\cdot$s$^{-1}$). We emphasise that CME predictions based on a 3D heliospheric model such as EUHFORIA are highly sensitive to the modelled background solar wind. For example, the fast stream that follows CME4 in Figure~\ref{fig:insitu}(a) was not captured in EUHFORIA's coronal model, resulting in a larger deceleration of the CME in interplanetary space and an arrival time that is a few hours later than observed. Moreover, since the semi-empirical coronal model uses synoptic magnetogram maps as input, it follows that the modelled solar wind originating from the backside of the Sun is subject to larger uncertainties, which likely resulted in discrepancies in the arrival times at STEREO-A. Nevertheless, our study showcases that 3D heliospheric models such as EUHFORIA are of high importance for providing a global context in the case of complex multiple-CME events. Finally, we remark that in this work we have employed a cone model for simulating the CMEs under analysis, thus we could not predict the arrival and the magnetic configuration of the ICME ejecta that were observed at 1~AU. Recent development efforts are under way to introduce magnetised ejecta into heliospheric models \citep[e.g.,][]{scolini2019,verbeke2019}, so that also the $B_{Z}$ component of the magnetic field and, consequently, the \emph{Dst} index could be estimated well in advance and compared with in-situ measurements.

In conclusion, we remark that moderate events play an important role in space weather research \citep[e.g.,][]{echer2013}. Forecasting moderate geomagnetic disturbances may be more difficult than extreme events because they are usually slower and signatures of Earth-directed components may be subtler, as shown in this study. These aspects highlight the importance of having a complete understanding of the whole heliospheric context when forecasting such events. They also emphasise the utility of having observations from multiple vantage points and combining various approaches to successfully capture the propagation and evolution of such ``problem events''. We demonstrated here the benefits of combining a detailed EUV, white-light, and radio analysis as well as heliospheric modelling. The ability to monitor and forecast solar transients and their geoeffectiveness may improve significantly through continuous observations away from the Sun--Earth line, especially from the solar poles or from Earth's Lagrange L5 point \citep[e.g.,][]{vourlidas2015,gibson2018a}.

\acknowledgments
E.P. acknowledges support from the PROBA2 Guest Investigator Programme and the Doctoral Programme in Particle Physics and Universe Sciences (PAPU) at the University of Helsinki. 
C.S. acknowledges support from the Research Foundation -- Flanders, FWO SB PhD fellowship 1S42817N.
A.N.Z. and L.R. thank the European Space Agency (ESA) and the Belgian Federal Science Policy Office (BELSPO) for their support in the framework of the PRODEX Programme.
S.W.G. and E.K.J.K. acknowledge Academy of Finland Project 1310445.
E.K.J.K. and J.P. acknowledge the European Research Council (ERC) under the European Union's Horizon 2020 Research and Innovation Programme Project SolMAG (grant agreement No 724391).
The authors thank an anonymous referee for their help in improving this manuscript.
EUHFORIA is developed as a joint effort between the University of Helsinki and KU Leuven. The validation of solar wind and CME modelling with EUHFORIA is being performed within the BRAIN-be project CCSOM (Constraining CMEs and Shocks by Observations and Modelling throughout the inner heliosphere; \url{http://www.sidc.be/ccsom}). For the simulation, the authors made use of the infrastructure of the VSC -- Flemish Supercomputer Center, funded by the Hercules foundation and the Flemish Government -- Department EWI.
The results presented in here have been achieved under the framework of the Finnish Centre of Excellence in Research of Sustainable Space (Academy of Finland grant number 1312390), which we gratefully acknowledge. 
We thank the WDC for Geomagnetism, Kyoto (\url{http://wdc.kugi.kyoto-u.ac.jp/wdc/Sec3.html}), and the geomagnetic observatories for their cooperation to make the final \emph{Dst} indices available. This paper uses data from the Heliospheric Shock Database (\url{http://ipshocks.fi}), generated and maintained at the University of Helsinki.
SWAP is a project of the Centre Spatial de Li\`ege and the Royal Observatory of Belgium funded by BELSPO. 
The HI instruments on STEREO were developed by a consortium that comprised the Rutherford Appleton Laboratory (UK), the University of Birmingham (UK), Centre Spatial de Li\`ege (CSL, Belgium) and the Naval Research Laboratory (NRL, USA). The STEREO/SECCHI project, of which HI is a part, is an international consortium led by NRL. We recognise the support of the UK Space Agency for funding STEREO/HI operations in the UK.

\facilities PROBA2 (SWAP); SDO (AIA, HMI); SOHO (LASCO); STEREO (SECCHI, IMPACT, PLASTIC, WAVES); \emph{Wind} (MFI, SWE, WAVES); NDA; NRH.
\software Python SunPy \citep{sunpy2015}, IDL SolarSoft \citep{freeland1998}, ESA JHelioviewer \citep{muller2017}.







\bibliography{bibliography.bib} 




\end{document}